%% file: main.tex
\pdfoutput=1
\documentclass[12pt,a4paper]{article}

\usepackage{ifthen} 
\newboolean{pdflatex}
\setboolean{pdflatex}{true} 

\newboolean{articletitles}
\setboolean{articletitles}{true} 

\newboolean{uprightparticles}
\setboolean{uprightparticles}{false} 

\def\paperauthors{LHCb collaboration} 
\def\paperasciititle{Measurement of the local and non-local components in B to Kmumu decays using LHCb data} 
\def\papertitle{Measurement of the local and nonlocal amplitudes in \decay{\Bp}{\Kp\mup\mun} decays} 
\def\paperkeywords{{High Energy Physics}, {LHCb}} 
\def\papercopyright{\the\year\ CERN for the benefit of the LHCb collaboration} 
\def\paperlicence{CC BY 4.0 licence}
\def\paperlicenceurl{https://creativecommons.org/licenses/by/4.0/}

\newif\ifEnableSectionTOCLinks
\EnableSectionTOCLinksfalse 

\input{preamble}
\input{paper-symbols}

\usepackage{longtable} 

\begin{document}

\renewcommand{\thefootnote}{\fnsymbol{footnote}}
\setcounter{footnote}{1}

\input{title-LHCb-PAPER}


\renewcommand{\thefootnote}{\arabic{footnote}}
\setcounter{footnote}{0}


\cleardoublepage


\pagestyle{plain} 
\setcounter{page}{1}
\pagenumbering{arabic}


\input{body}

\input{acknowledgements}


\addcontentsline{toc}{section}{References}
\bibliographystyle{LHCb}
\bibliography{main,standard,LHCb-PAPER,LHCb-CONF,LHCb-DP,LHCb-TDR}

\newpage
\input{Authorship_LHCb-PAPER-2025-055}

\end{document}

%% file: preamble.tex
\usepackage[top=1in, bottom=1.25in, left=1in, right=1in]{geometry}

\usepackage{microtype}
\usepackage{lineno}  
\usepackage{xspace} 
\usepackage{caption} 

\usepackage{graphicx}  
\usepackage{color}
\usepackage{colortbl}
\graphicspath{{./figs/}} 

\usepackage{amsmath} 
\usepackage{amssymb}
\usepackage{amsfonts}
\usepackage{upgreek} 
\usepackage{booktabs}
\newcommand*\patchAmsMathEnvironmentForLineno[1]{%
\expandafter\let\csname old#1\expandafter\endcsname\csname #1\endcsname
\expandafter\let\csname oldend#1\expandafter\endcsname\csname
end#1\endcsname
 \renewenvironment{#1}%
   {\linenomath\csname old#1\endcsname}%
   {\csname oldend#1\endcsname\endlinenomath}%
}
\newcommand*\patchBothAmsMathEnvironmentsForLineno[1]{%
  \patchAmsMathEnvironmentForLineno{#1}%
  \patchAmsMathEnvironmentForLineno{#1*}%
}
\AtBeginDocument{%
\patchBothAmsMathEnvironmentsForLineno{equation}%
\patchBothAmsMathEnvironmentsForLineno{align}%
\patchBothAmsMathEnvironmentsForLineno{flalign}%
\patchBothAmsMathEnvironmentsForLineno{alignat}%
\patchBothAmsMathEnvironmentsForLineno{gather}%
\patchBothAmsMathEnvironmentsForLineno{multline}%
\patchBothAmsMathEnvironmentsForLineno{eqnarray}%
}

\usepackage[pdftex]{hyperref}
\usepackage{hyperxmp}
\hypersetup{
  pdfauthor     = {\paperauthors},
  pdftitle      = {\paperasciititle},
  pdfkeywords   = {\paperkeywords},
  pdfcopyright  = {Copyright (C) \papercopyright},
  pdflicenseurl = {\paperlicenceurl},
  colorlinks    = true,
  urlcolor      = blue,
  linkcolor     = blue,
  citecolor     = red
}

\usepackage[bottom,flushmargin,hang,multiple]{footmisc}

\usepackage[all]{hypcap} 

\input{lhcb-symbols-def} 

\hypersetup{
  colorlinks   = true, 
  urlcolor     = blue, 
  linkcolor    = blue, 
  citecolor    = red   
}

\ifEnableSectionTOCLinks
    \usepackage[explicit]{titlesec} 
    
    \let\oldcontentsline\contentsline
    \renewcommand

    \titleformat{\section}{\normalfont\Large\bf}{\hyperlink{tocsection.\thesection}{{\thesection} \parbox[t]{\dimexpr\textwidth-1pc}{#1}}}{1pc}{}

    \titleformat{\subsection}{\normalfont\bf}{\hyperlink{tocsubsection.\thesubsection}{{\thesubsection} \parbox[t]{\dimexpr\textwidth-1pc}{#1}}}{1pc}{}

    \titleformat{name=\section,numberless}[display]{}{}{0pt}{\normalfont\Huge\bfseries #1}
\fi

\usepackage{cite} 
\usepackage{mciteplus}

%% file: lhcb-symbols-def.tex
\usepackage{xspace} 
\usepackage{upgreek}

\def\lhcb   {\mbox{LHCb}\xspace}

\def\MagUp {\mbox{\em Mag\kern -0.05em Up}\xspace}

\ifthenelse{\boolean{uprightparticles}}%
{

 \def\Pmu         {\ensuremath{\upmu}\xspace}

 \def\Ppi         {\ensuremath{\uppi}\xspace}

 \def\Ppsi        {\ensuremath{\uppsi}\xspace}

 \def\PDelta      {\ensuremath{\Delta}\xspace}                 
 \def\PXi         {\ensuremath{\Xi}\xspace}                 
 \def\PLambda     {\ensuremath{\Lambda}\xspace}                 
 \def\PSigma      {\ensuremath{\Sigma}\xspace}                 
 \def\POmega      {\ensuremath{\Omega}\xspace}                 
 \def\PUpsilon    {\ensuremath{\Upsilon}\xspace}
 \let\oldPi\Pi
 \def\PPi         {\ensuremath{\oldPi}\xspace}

 \def\PB      {\ensuremath{\mathrm{B}}\xspace}                 
 \def\PD      {\ensuremath{\mathrm{D}}\xspace}                 
 \def\PJ      {\ensuremath{\mathrm{J}}\xspace}                 
 \def\PK      {\ensuremath{\mathrm{K}}\xspace}                 
 \def\Pb      {\ensuremath{\mathrm{b}}\xspace}                 
 \def\Pc      {\ensuremath{\mathrm{c}}\xspace}

 \def\Pp      {\ensuremath{\mathrm{p}}\xspace}                 

 \def\Ps      {\ensuremath{\mathrm{s}}\xspace}                 
 \def\Pt      {\ensuremath{\mathrm{t}}\xspace}                 

 \def\thebaroffset{0.0em}
}
{

 \def\Pmu         {\ensuremath{\mu}\xspace}

 \def\Ppi         {\ensuremath{\pi}\xspace}

 \def\Ppsi        {\ensuremath{\psi}\xspace}                 
                  
 \mathchardef\PDelta="7101
 \mathchardef\PXi="7104
 \mathchardef\PLambda="7103
 \mathchardef\PSigma="7106
 \mathchardef\POmega="710A
 \mathchardef\PUpsilon="7107
 \mathchardef\PPi="7105
 \def\PB      {\ensuremath{B}\xspace}                 
 \def\PD      {\ensuremath{D}\xspace}                 
 \def\PJ      {\ensuremath{J}\xspace}                 
 \def\PK      {\ensuremath{K}\xspace}                 
 \def\Pb      {\ensuremath{b}\xspace}                 
 \def\Pc      {\ensuremath{c}\xspace}

 \def\Pp      {\ensuremath{p}\xspace}                 

 \def\Ps      {\ensuremath{s}\xspace}                 
 \def\Pt      {\ensuremath{t}\xspace}                 

 \def\thebaroffset{0.18em}
}
\newcommand{\offsetoverline}[2][\thebaroffset]{\kern #1\overline{\kern -#1 #2}}%

\makeatletter
\ifcase \@ptsize \relax
  \newcommand{\miniscule}{\@setfontsize\miniscule{4}{5}}
\or
  \newcommand{\miniscule}{\@setfontsize\miniscule{5}{6}}
\or
  \newcommand{\miniscule}{\@setfontsize\miniscule{5}{6}}
\fi
\makeatother

\DeclareRobustCommand{\optbar}[1]{\shortstack{{\miniscule (\rule[.5ex]{1.25em}{.18mm})}
  \\ [-.7ex] $#1$}}

\def\muon       {{\ensuremath{\Pmu}}\xspace}
\def\mup        {{\ensuremath{\Pmu^+}}\xspace}
\def\mun        {{\ensuremath{\Pmu^-}}\xspace} 

\def\mumu       {{\ensuremath{\Pmu^+\Pmu^-}}\xspace}

\def\squark    {{\ensuremath{\Ps}}\xspace}

\def\cquark    {{\ensuremath{\Pc}}\xspace}
\def\cquarkbar {{\ensuremath{\overline \cquark}}\xspace}
\def\ccbar     {{\ensuremath{\cquark\cquarkbar}}\xspace}
\def\bquark    {{\ensuremath{\Pb}}\xspace}

\def\tquark    {{\ensuremath{\Pt}}\xspace}

\def\pion   {{\ensuremath{\Ppi}}\xspace}

\def\pip    {{\ensuremath{\pion^+}}\xspace}
\def\pim    {{\ensuremath{\pion^-}}\xspace}

\def\kaon    {{\ensuremath{\PK}}\xspace}
\def\Kbar    {{\ensuremath{\offsetoverline{\PK}}}\xspace}

\def\KorKbar {\kern \thebaroffset\optbar{\kern -\thebaroffset \PK}{}\xspace}
\def\Kz      {{\ensuremath{\kaon^0}}\xspace}
\def\Kzb     {{\ensuremath{\Kbar{}^0}}\xspace}
\def\Kp      {{\ensuremath{\kaon^+}}\xspace}
\def\Km      {{\ensuremath{\kaon^-}}\xspace}

\def\Kstarz  {{\ensuremath{\kaon^{*}(892)^{0}}}\xspace}

\def\Dbar    {{\ensuremath{\offsetoverline{\PD}}}\xspace}
\def\D       {{\ensuremath{\PD}}\xspace}
\def\Db      {{\ensuremath{\Dbar}}\xspace}
\def\DorDbar {\kern \thebaroffset\optbar{\kern -\thebaroffset \PD}\xspace}

\def\Dzb     {{\ensuremath{\Dbar{}^0}}\xspace}
\def\Dp      {{\ensuremath{\D^+}}\xspace}
\def\Dm      {{\ensuremath{\D^-}}\xspace}

\def\DpDm    {\ensuremath{\Dp {\kern -0.16em \Dm}}\xspace}
\def\Dstar   {{\ensuremath{\D^*}}\xspace}
\def\DorDstar   {{\ensuremath{\D^{(*)}}\xspace}}
\def\Dstarb  {{\ensuremath{\Dbar{}^*}}\xspace}
\def\DorDstarb  {{\ensuremath{\Dbar{}^{(*)}}\xspace}}
\def\Dstarb  {{\ensuremath{\Dbar{}^*}}\xspace}

\def\B       {{\ensuremath{\PB}}\xspace}

\def\BorBbar {\kern \thebaroffset\optbar{\kern -\thebaroffset \PB}\xspace}

\def\Bd      {{\ensuremath{\B^0}}\xspace}

\def\BdorBdbar {\kern \thebaroffset\optbar{\kern -\thebaroffset \Bd}\xspace}
\def\Bu      {{\ensuremath{\B^+}}\xspace}
\def\Bub     {{\ensuremath{\B^-}}\xspace}
\def\Bp      {{\ensuremath{\Bu}}\xspace}
\def\Bm      {{\ensuremath{\Bub}}\xspace}

\def\Bs      {{\ensuremath{\B^0_\squark}}\xspace}

\def\BsorBsbar {\kern \thebaroffset\optbar{\kern -\thebaroffset \Bs}\xspace}

\def\jpsi     {{\ensuremath{{\PJ\mskip -3mu/\mskip -2mu\Ppsi}}}\xspace}
\def\psitwos  {{\ensuremath{\Ppsi{(2S)}}}\xspace}

\def\Y#1S{\ensuremath{\PUpsilon{(#1S)}}\xspace}

\def\proton      {{\ensuremath{\Pp}}\xspace}

\def\LorLbar     {\kern \thebaroffset\optbar{\kern -\thebaroffset \PLambda}\xspace}

\def\BF         {{\ensuremath{\mathcal{B}}}\xspace}

\newcommand{\decay}[2]{\ensuremath{#1\!\to #2}\xspace} 

\def\to                 {\ensuremath{\rightarrow}\xspace}

\def\qsq       {{\ensuremath{q^2}}\xspace}

\def\CP                {{\ensuremath{C\!P}}\xspace}

\def\Vtb  {{\ensuremath{V_{\tquark\bquark}^{\phantom{\ast}}}}\xspace}

\def\Vtss  {{\ensuremath{V_{\tquark\squark}^\ast}}\xspace}

\def\BdToKstmm    {\decay{\Bd}{\Kstarz\mup\mun}}

\def\bsll     {\decay{\bquark}{\squark \ell^+ \ell^-}}

\def\AT#1     {\ensuremath{A_{\mathrm{T}}^{#1}}\xspace}           

\def\ctl       {\ensuremath{\cos{\theta_\ell}}\xspace}

\def\C#1      {\ensuremath{\mathcal{C}_{#1}}\xspace}                       
\def\Cp#1     {\ensuremath{\mathcal{C}_{#1}^{'}}\xspace}                    
\def\Ceff#1   {\ensuremath{\mathcal{C}_{#1}^{\mathrm{(eff)}}}\xspace}        
\def\Cpeff#1  {\ensuremath{\mathcal{C}_{#1}^{'\mathrm{(eff)}}}\xspace}       
\def\Ope#1    {\ensuremath{\mathcal{O}_{#1}}\xspace}                       
\def\Opep#1   {\ensuremath{\mathcal{O}_{#1}^{'}}\xspace}                    

\newcommand{\nospaceunit}[1]{\ensuremath{\text{#1}}}       
\newcommand{\aunit}[1]{\ensuremath{\text{\,#1}}}       

\newcommand{\tev}{\aunit{Te\kern -0.1em V}\xspace}
\newcommand{\gev}{\aunit{Ge\kern -0.1em V}\xspace}
\newcommand{\mev}{\aunit{Me\kern -0.1em V}\xspace}
\newcommand{\kev}{\aunit{ke\kern -0.1em V}\xspace}
\newcommand{\ev}{\aunit{e\kern -0.1em V}\xspace}
 
\newcommand{\mevc}{\ensuremath{\aunit{Me\kern -0.1em V\!/}c}\xspace}
\newcommand{\gevc}{\ensuremath{\aunit{Ge\kern -0.1em V\!/}c}\xspace}
\newcommand{\mevcc}{\ensuremath{\aunit{Me\kern -0.1em V\!/}c^2}\xspace}
\newcommand{\gevcc}{\ensuremath{\aunit{Ge\kern -0.1em V\!/}c^2}\xspace}
\newcommand{\gevgevcccc}{\ensuremath{\gev^2\!/c^4}\xspace} 

\def\mum  {\ensuremath{\,\upmu\nospaceunit{m}}\xspace}

\def\fb   {\ensuremath{\aunit{fb}}\xspace}
\def\invfb   {\ensuremath{\fb^{-1}}\xspace}

\def\deriv {\ensuremath{\mathrm{d}}}

\def\gsim{{~\raise.15em\hbox{$>$}\kern-.85em
          \lower.35em\hbox{$\sim$}~}\xspace}
\def\lsim{{~\raise.15em\hbox{$<$}\kern-.85em
          \lower.35em\hbox{$\sim$}~}\xspace}

\def\pt         {\ensuremath{p_{\mathrm{T}}}\xspace}

\def\ptot       {\ensuremath{p}\xspace}

\def\evtgen     {\mbox{\textsc{EvtGen}}\xspace}

\def\geant      {\mbox{\textsc{Geant4}}\xspace}

\def\photos     {\mbox{\textsc{Photos}}\xspace}

\def\pythia     {\mbox{\textsc{Pythia}}\xspace}

\def\tell1  {TELL1\xspace}
\def\ukl1   {UKL1\xspace}

\newcommand{\eg}{\mbox{\itshape e.g.}\xspace}

\newcommand{\lhcborcid}[1]{\href{https://orcid.org/#1}{\hspace*{0.1em}\raisebox{-0.45ex}{\includegraphics[width=1em]{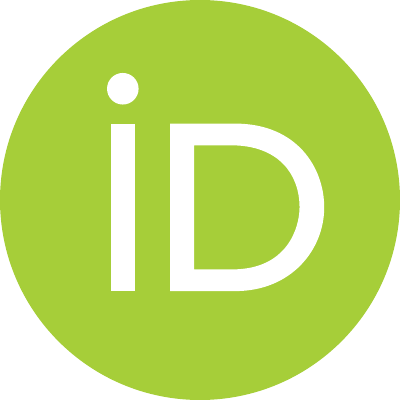}}}}


%% file: paper-symbols.tex
\def\bsmumu {\decay{\bquark}{\squark\mup\mun}}

\def\BKmumu {\decay{\Bp}{\Kp\mup\mun}}

\newcommand{\myC}[1]{\ensuremath{C_{#1}}\xspace}             
\newcommand{\myCp}[1]     {\ensuremath{C_{#1}^{'}}\xspace}                    
\newcommand{\myCeff}[1]  {\ensuremath{C_{#1}^{\mathrm{eff}}}\xspace}        
\newcommand{\myOpe}[1]   {\ensuremath{\mathcal{O}_{#1}}\xspace}                       

\def\hameff {\ensuremath{\mathcal{H}^{\rm eff}}\xspace}
\def\Gfermi {\ensuremath{G_F}\xspace }

\def\mmumu {\ensuremath{m_{\muon\muon}}\xspace}
\def\mKmumu {\ensuremath{m_{\kaon\muon\muon}}\xspace}

\def\psitss {{\ensuremath{\psi{(3770)}}}\xspace}
\def\psifzfz {{\ensuremath{\psi{(4040)}}}\xspace}
\def\psifosz {{\ensuremath{\psi{(4160)}}}\xspace}
\def\psiffof {{\ensuremath{\psi{(4415)}}}\xspace}

\def\rhossz {\ensuremath{\rho (770)}\xspace}
\def\omegaset {\ensuremath{\omega (782)}\xspace}
\def\phioztz {\ensuremath{\phi (1020)}\xspace}

\def\lessthan  {\ensuremath{<}\xspace}
\def\greaterthan  {\ensuremath{>}\xspace}

\def\Spm {\ensuremath{\pm}\xspace}

\newcommand{\phase}[1]{\ensuremath{\delta_{#1}}\xspace}
\newcommand{\magni}[1]{\ensuremath{\eta_{#1}}\xspace}

\newcommand{\smhpqcdsig}{\ensuremath{4.0\sigma}\xspace}

\newcommand{\smfnalsig}{\ensuremath{1.6\sigma}\xspace}

%% file: title-LHCb-PAPER.tex

\begin{titlepage}
\pagenumbering{roman}

\vspace*{-1.5cm}
\centerline{\large EUROPEAN ORGANIZATION FOR NUCLEAR RESEARCH (CERN)}
\vspace*{1.5cm}
\noindent
\begin{tabular*}{\linewidth}{lc@{\extracolsep{\fill}}r@{\extracolsep{0pt}}}
\ifthenelse{\boolean{pdflatex}}
{\vspace*{-1.5cm}\mbox{\!\!\!\includegraphics[width=.14\textwidth]{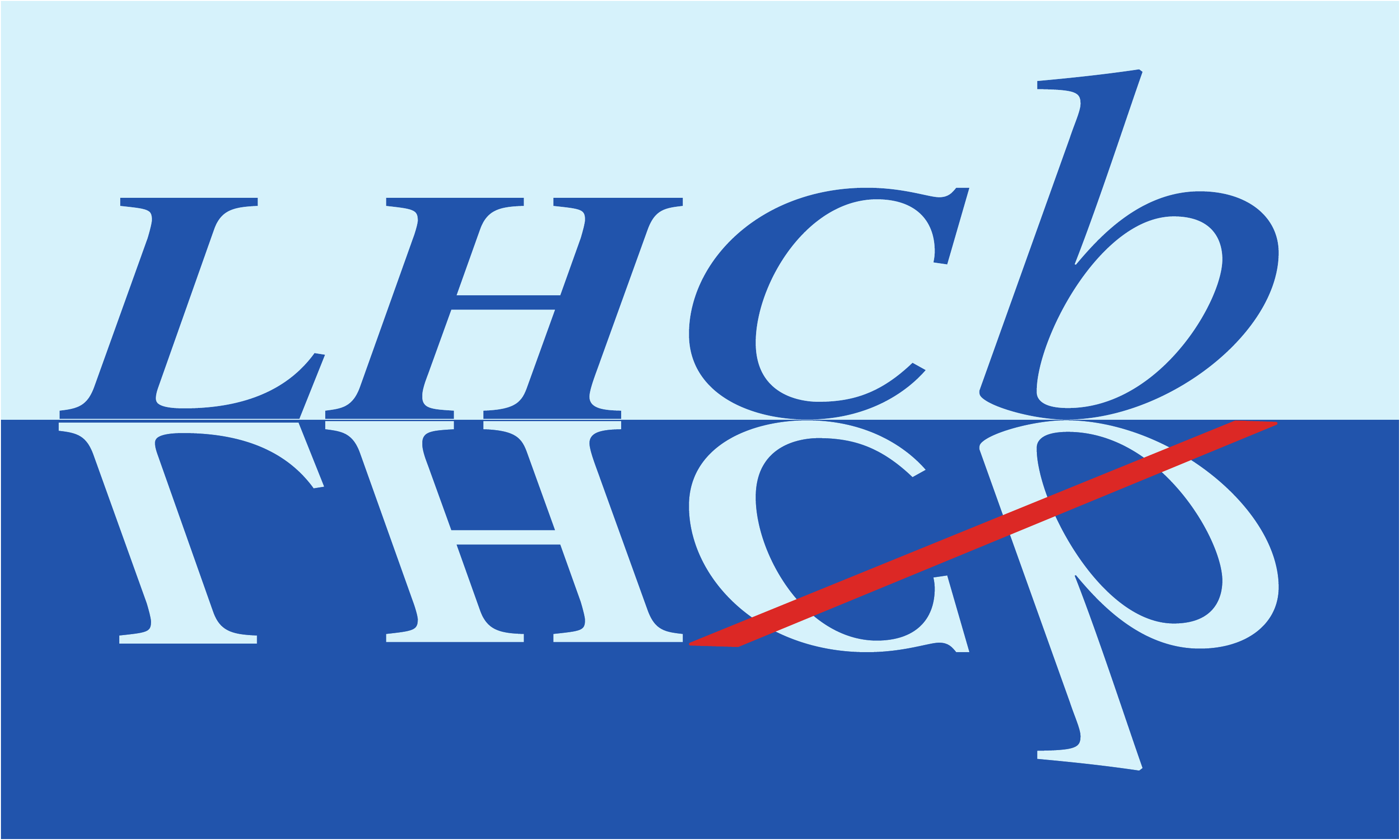}} & &}%
{\vspace*{-1.2cm}\mbox{\!\!\!\includegraphics[width=.12\textwidth]{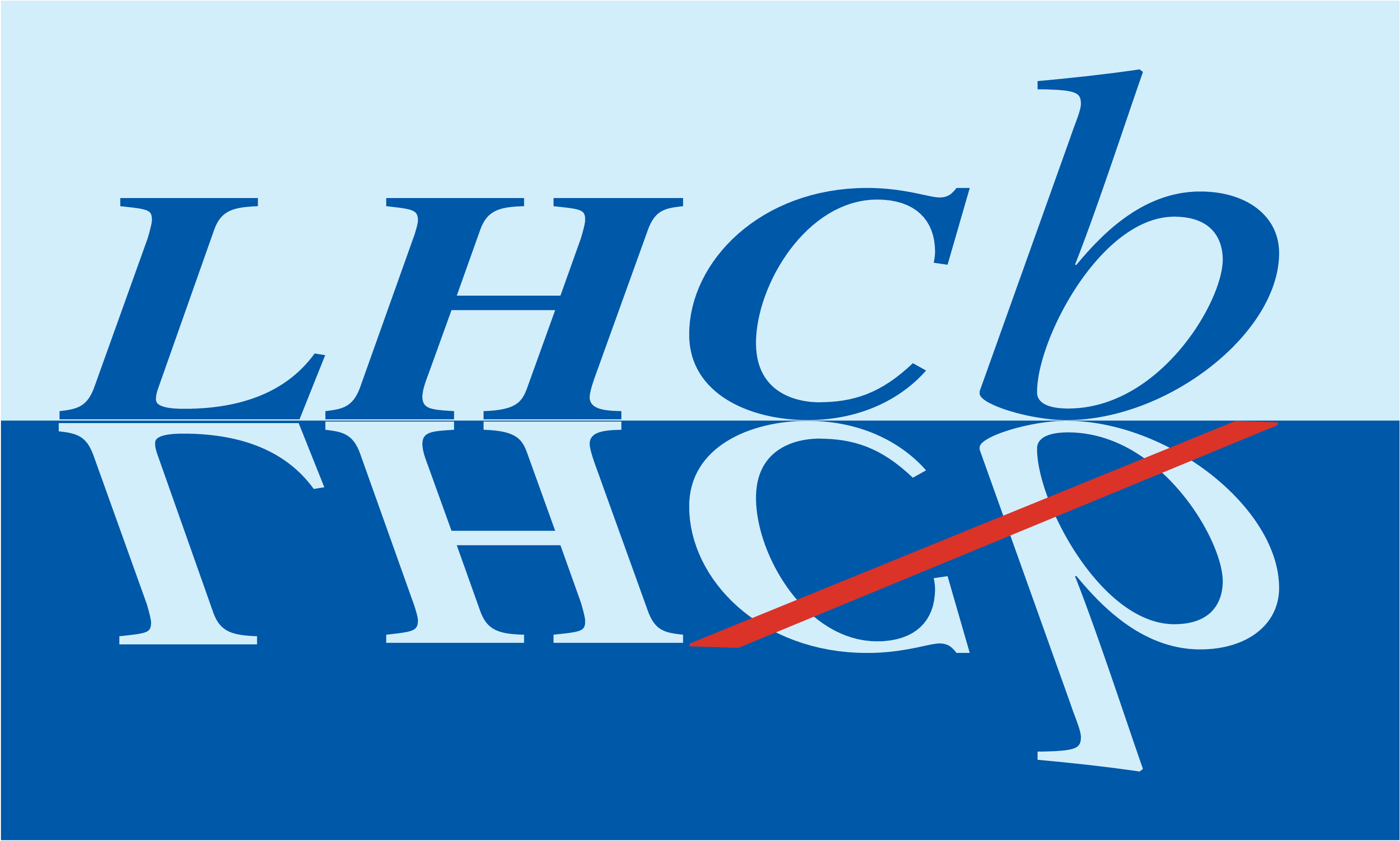}} & &}%
\\
 & & CERN-EP-2026-018 \\  
 & & LHCb-PAPER-2025-055  \\  
 & & 4 June 2026 \\ 
 & & \\
\end{tabular*}

\vspace*{4.0cm}

{\normalfont\bfseries\boldmath\huge
\begin{center}
  \papertitle 
\end{center}
}

\vspace*{0.8cm}

\begin{center}
\paperauthors\footnote{Authors are listed at the end of this paper.}
\end{center}

\vspace{\fill}

\begin{abstract}
  \noindent
  This paper presents a thorough study of the local and nonlocal amplitudes in \BKmumu transitions through an amplitude analysis of the dimuon mass spectrum of the decay. The analysis is based on $pp$ collision data corresponding to an integrated luminosity of 8.4\invfb collected by the LHCb experiment during Run1 and Run2. This measurement employs a model that describes both one-particle and two-particle nonlocal amplitudes across the entirety of the dimuon mass spectrum, enabling the determination 
of both short- and long-distance contributions to the decay. The compatibility of the Wilson coefficient combinations $C_9+C_9'$ and $C_{10}+C_{10}'$ with the Standard Model prediction is found to vary between  $1.6\,\sigma$ and $4\,\sigma$, depending on the choice of local form factors.

\end{abstract}

\vspace*{2.0cm}

\begin{center}
  Published in JHEP 08 (2026) 155
\end{center}

\vspace{\fill}

{\footnotesize 
\centerline{\copyright~\papercopyright. \href{\paperlicenceurl}{\paperlicence}.}}
\vspace*{2mm}

\end{titlepage}


\newpage
\setcounter{page}{2}
\mbox{~}
%
%
%
%

%% file: body.tex
\section{Introduction}
\label{sec:Introduction}

Over the past decade, a broad set of measurements of \bsmumu transitions has revealed a consistent pattern of tensions with Standard Model (SM) predictions, attracting considerable interest from both the theoretical and experimental communities~\cite{Mahmoudi:2024zna,Andersson:2024nam}. In this context, the decay \BKmumu plays a central role as it is both experimentally and theoretically simpler compared to other \bsmumu transitions, since only a single charged pseudoscalar hadron is present in the final state.\footnote{The inclusion of charge-conjugate processes is implied throughout, unless otherwise specified.} This paper presents a model-dependent analysis of the long- and short-distance contributions in \BKmumu decays across the reconstructed dimuon mass range $300<m_{\mu\mu}<4700\mevcc$ using a data sample corresponding to an integrated luminosity of 8.4\invfb of \proton\proton collisions collected during Run~1 and Run~2 of the LHCb experiment. The measurements presented here supersede those in Ref.~\cite{LHCb-PAPER-2016-045} by employing a more complete description of the long-distance amplitudes and including Run~2 data.

The \bsll transitions are studied in the framework of weak effective theories where the dimension-six effective Hamiltonian is defined as a sum of operators (\myOpe{i}) with Wilson  coefficients \myC{i},
\begin{equation}
    \hameff = -\frac{4\Gfermi}{\sqrt{2}}\Vtb\Vtss \sum_{i} \myC{i}^{(')} (\mu) \myOpe{i}^{(')} (\mu), 
    \label{eq:eff_hamil_btosll}
\end{equation}
where $\Gfermi$ is the Fermi constant, $V^{(*)}_{tq}$ are the Cabibbo--Kobayashi--Maskawa (CKM) matrix elements, and $\mu$ defines the
energy scale.
The Wilson coefficients, $\myC{i}^{(')}$, encode the heavy short-distance physics, while the operators, $\myOpe{i}^{(')}$, account for the long-distance contributions. By measuring the Wilson coefficients, the presence of short-distance new physics can be established. 

The local contributions to the decay rate correspond to the SM electroweak penguin contribution and new-physics amplitudes, which mediate the decay. The nonlocal contributions are generated through four-quark operators that interact electromagnetically with a $\mu^+\mu^-$ pair. As described in Ref.~\cite{Cornella:2020aoq}, these nonlocal amplitudes can be modelled through the sum of one-particle (1p) $J^{PC}=1^{--}$ charmonium and light-quark states, and two-particle (2p) amplitudes comprising $B^+\to K^+(\DorDstar\DorDstarb\to\mu^+\mu^-)$ processes, where $\DorDstar$ denotes a ground-state or excited charm meson.

Global analyses of differential branching fraction measurements and angular distributions of \mbox{\BKmumu}, \mbox{\BdToKstmm} and other \bsmumu decays performed in bins of the squared invariant mass of the dimuon system (\qsq), exhibit tensions with SM predictions~\cite{LHCb-PAPER-2025-041,LHCb-PAPER-2021-014,LHCb-PAPER-2020-041,LHCb-PAPER-2014-006,CMS:2024syx,ATLAS:2018gqc,Belle:2016fev, Hurth:2025vfx, Bordone:2024hui,Alguero:2023jeh,Gubernari:2022hxn} consistent with the presence of new physics in the Wilson coefficient \myC{9}. However, there is considerable debate regarding the possibility of underestimated nonlocal contributions in the SM predictions of \bsmumu observables that can mimic new physics~\cite{Khodjamirian:2010vf,Korchin:2012kz,Jager:2012uw, Descotes-Genon:2013wba,Lyon:2014hpa,Jager:2014rwa,Ciuchini:2015qxb,Gubernari:2020eft,Ciuchini:2022wbq,Gubernari:2022hxn,PhysRevD.111.093007,isidori2025charmrescatteringb0tok0barellell}. 

Recently, the LHCb collaboration performed amplitude analyses of the \mbox{$\BdToKstmm$} decay using different models to describe the nonlocal amplitude across the decay phase space~\cite{LHCb-PAPER-2024-011, LHCb-PAPER-2023-033}. These measurements resulted in a data-driven determination of the dominant hadronic contributions while confirming the tension in \myC{9}, albeit with reduced statistical precision. 

The paper is structured as follows: a detailed description of the nonlocal model is given in Sec.~\ref{sec:diff_rate}. 
In Sec.~\ref{sec:det_sims} the LHCb detector and simulation procedure are introduced, and in Sec.~\ref{sec:det_sel} the candidate selection procedure is discussed. 
The selection efficiency, resolution and background models are discussed in Sec.~\ref{sec:decay_rate_in_data}, followed by a description of the sources of systematic uncertainty and the method adopted to account for them in Sec.~\ref{sec:syst}.
The fit to the dimuon spectrum and the systematic effects are presented and the results discussed in Sec.~\ref{sec:result_and_discussion}, followed by the conclusion in Sec.~\ref{sec:conclusion}.

\section{Differential decay rate}
\label{sec:diff_rate}
Following the model described in Ref.~\cite{Cornella:2020aoq}, the \CP-averaged differential decay rate of \mbox{\BKmumu} decays is given by
\begin{align}
    \frac{\deriv\Gamma_\muon}{\deriv q^2} = & \frac{G_F^2\alpha^2|\Vtb\Vtss|^2}{128\pi^5} |\mathbf{k}|\beta_+
         \left\{ \vphantom{\left|2\Ceff7 \frac{m_b+m_s}{M_B+M_K} f_T(q^2)\right|^2}
         \frac{2}{3}|\mathbf{k}|^2\beta^2_+ \left|\myC{A} f_+(q^2)\right|^2 \right.
     +
         \frac{m_{\muon}^2 (M_B^2-M_K^2)^2 }{q^2 M_B^2}
         \left|\myC{A} f_0(q^2)\right|^2
     \nonumber \\ & { } + \left.
         |\mathbf{k}|^2 \left[1 - \frac{1}{3}\beta^2_+ \right]
         \left| \myC{V} f_+(q^2) + 2(\myCeff{7}+\myC{7}') \frac{m_b+m_s}{M_B+M_K} f_T(q^2) \right|^2
         \right\}\ ,
 \label{eq:diff_bran_frac_theory_ch}
\end{align}
where $M_K$ and $M_B$ are the masses of the kaon and \B meson, $m_b$, $m_s$ and $m_{\muon}$ are the masses of the \bquark quark, \squark quark and the muon, respectively. 
The factor $|\mathbf{k}|$  is the magnitude of the kaon momentum in the rest frame of the \B meson and $\beta^2_+\equiv 1-4m^2_{\muon}/\qsq$.
The parameter $\alpha$ is the QED fine structure constant. 
The parameters $f_{0,+,T}$ are the scalar, vector and tensor $\decay{B}{K}$ form factors. This analysis employs lattice QCD computations for these quantities.

The Wilson coefficients $\myC{V,A}$  are defined as $\myC{V,A}\equiv \myC{9,10}+\myCp{9,10}$, where $\myC{9}$, $\myC{10}$ correspond to the coupling strength of the vector and axial-vector current operators \myOpe{9} and \myOpe{10}. The Wilson coefficients $\myCp{9}$, $\myCp{10}$ correspond to the right-handed counterparts of operators $\myOpe{9}'$ and $\myOpe{10}'$. 
The dependence of Eq.~\ref{eq:diff_bran_frac_theory_ch} on $\myC{V,A}$ makes a fit to the differential decay rate of \BKmumu transitions unable to distinguish between the left- and right-handed counterparts of \myC{9} and \myC{10}. 
For the measurements presented in this paper, \myCeff{7} is set to the SM value given in Ref.~\cite{Bordone:2024hui}, $\myC{7}'$ is set to zero, and all Wilson coefficients are assumed to be real. 
The values of all the constants used in Eq.~\ref{eq:diff_bran_frac_theory_ch} are summarised in Table \ref{tab:constants}.
\begin{table}[!t]
    \centering
    \caption{The values for the constants used in the measurement presented in this paper, taken from Refs.~\cite{Bordone:2024hui,PDG2024}. }
    \label{tab:constants}
    \begin{tabular}{cc}
        \toprule
        Constant & Value \\
        \midrule
        $G_F$ & 1.1745$\times 10^{-5}\gev^{-2}$ \\
        $|\Vtb\Vtss|$ & 0.04185\\
        $M_B$ & 5279.26\mevcc\\
        $M_K$ & 493.677\mevcc\\
        $m_b$ & 4870\mevcc\\
        $m_s$ & 95\mevcc\\
        $m_\mu$ & 105.658\mevcc\\
        $\myCeff{7}$ & $-0.450$\\
        $\alpha(m_b)$ & 1/133\\
        \bottomrule
    \end{tabular}
\end{table}

The $\qsq$-dependent nonlocal amplitudes are included in the decay rate through the redefinition of
\begin{equation}
    \myC{9} \rightarrow \myCeff{9} \equiv \myC{9} + Y^{\rm 1p}_{\mathrm{light}} (\qsq) + Y^{\rm 1p+2p}_{\ccbar} (\qsq),
\end{equation}
where $Y^{\rm 1p}_{\mathrm{light}}$ and $Y^{\rm 1p+2p}_{\ccbar}$ are the total amplitudes of the light one-particle states and the sum of one-particle and two-particle \ccbar states, respectively.
Following Ref.~\cite{Cornella:2020aoq}, the functions $Y^{\rm 1p}_{\mathrm{light}}$ and $Y^{\rm 1p+2p}_{\ccbar}$ are written in terms of hadronic dispersion relations.

The one- and two-particle \ccbar amplitudes are given by the following expression
\begin{equation}
    Y^{\rm 1p+2p}_\ccbar(\qsq) = Y^{0}_{\ccbar}(q^{2}_{0}) + \Delta Y^{\rm 1p}_\ccbar(\qsq) + \Delta Y^{\rm 2p}_\ccbar(\qsq),
    \label{eq:non_local_cc_breakdown}
\end{equation}
with $Y^{0}_{\ccbar}(q^{2}_{0})$ representing the subtraction constant of the dispersion relation.
In this analysis, the subtraction point is chosen to be $q^2_0 = 0$, following Ref.~\cite{Bordone:2024hui}, with \mbox{$Y^{0}_{\ccbar}(0) = 0.230\pm0.065$}. The one-particle \ccbar contribution is modelled as
\begin{align}
    \Delta Y^{\rm 1p}_\ccbar (\qsq) &= \sum_{j} \eta_j e^{i\delta_j} \frac{(\qsq-q^2_0)}{(m^2_j-q^2_0)} A^{\rm res}_j(\qsq) \label{eq:cc_1p_contrib},
\end{align}
where $\eta_j$ and $\delta_j$ are the magnitude and phase of the $j^{\rm th}$ resonant contribution  relative to the local \BKmumu amplitude. 
For the \jpsi, \psitwos, \psifzfz, \psifosz and \psiffof resonances, the factor $A^{\rm res}_j(\qsq)$ is given by a relativistic Breit--Wigner function as shown below   
\begin{equation}
A^{\rm res}_j(\qsq) =  \frac{m_j\Gamma_{0j}}{(m_j^2 - \qsq) - im_j\Gamma_j(\qsq)},
\label{eq:relbw}
\end{equation}
where $m_j$ and $\Gamma_{0j}$ represent the mass and natural width of the resonance and the running width $\Gamma_j(\qsq)$ is given by 
\begin{align}
    &\Gamma_j(\qsq) = \frac{p(m_\mu,\qsq)}{p(m_\mu,m^2_j)} \frac{m_j}{\sqrt{\qsq}} \Gamma_{0j}, \label{eq:generic_running_width}
\end{align}
with
\begin{align}
    p(m,x) &= \frac{\sqrt{x}}{2} \sqrt{1-\frac{4m^{2}}{x}} \label{eq:momentum_eval_atX}.
\end{align}

\noindent For the \psitss resonance $A^{\rm res}_j(\qsq)$ takes the form of a Flatt\'{e} function~\cite{Flatte:1976xu} to account for the open-charm threshold at $2m_D$. Therefore, in this case, the running width of the \psitss state is given by
\begin{align}
    &\Gamma_j(\qsq) = \frac{p(m_\mu,\qsq)}{p(m_\mu,m^2_j)} \frac{m_j}{\sqrt{\qsq}} \Gamma_{\psitss,1} + \frac{p(m_D,\qsq)}{p(m_D,m^2_j)} \frac{m_j}{\sqrt{\qsq}} \Gamma_{\psitss,2}.
\end{align}
The two widths $\Gamma_{\psitss,1}$ and $\Gamma_{\psitss,2}$ are taken from Ref.~\cite{PDG2024} and correspond to the partial widths of \psitss decaying to final states below and above the open-charm threshold. 
The two-particle charm contributions are given by,
\begin{equation}
    \Delta Y^{\rm 2p}_\ccbar (\qsq) = \sum_{j}  \eta_{\rm 2p}^{j} e^{i\delta_{\rm 2p}^{j}} A^{\rm 2p}_j (\qsq) \label{eq:y2p_amplitude} 
\end{equation}
with
\begin{equation}
    A^{\rm 2p}_{j} (\qsq) = \frac{(\qsq-q^2_0)}{\pi} \int_{q^2_{min}}^{\infty} \deriv s \frac{\hat{\rho}_j (s)}{(s-q^2_0)(s-\qsq - i\epsilon)},
    \label{eq:y2p_sum1} 
\end{equation} 
where the sum in Eq.~\ref{eq:y2p_amplitude} runs over the $\D\Dbar$, $\D\Dstarb$ and $\Dstar\Dstarb$ contributions, and the normalised spectral densities of the two-particle states are modelled by the two-body phase space $\hat{\rho}_j$.
The lower threshold of the integral $q^2_{\rm min}$ is taken to be just below the $D\Db$ threshold. The parameters $\eta_{\rm 2p}^{j}$ and $\delta_{\rm 2p}^{j}$ are the magnitude and phase of each of the two-particle states and are measured from data. Following the approach of Ref.~\cite{Cornella:2020aoq}, the conservation of angular momentum and parity in the two-particle production dictate that the two-particle states are produced in a P-wave configuration. In this case, the integral in Eq.~\ref{eq:y2p_sum1} can be approximated to give

\begin{equation}
    \Delta Y^{\rm 2p}_{\ccbar} (\qsq) = \eta_{\rm 2p} e^{i\delta_{\rm 2p}} \sum_{j} h_P\left(m_j, \qsq \right)
    \label{eq:2p_state_amp_approx} 
\end{equation}
with
\begin{equation}
    h_P \left(m,  q^2 \right) = \frac{2}{3}+ \left( 1 -\frac{4 m^2}{q^2} \right)  h_S \left(m,  q^2 \right),\qquad
    h_S \left(m,  q^2 \right) = 2-  G\left( 1 -\frac{4 m^2}{q^2} \right), 
    \label{eq:hs_hp_dispersion}
\end{equation}
where 
\begin{align}
    G(y) =  \sqrt{|y|} \left\{ \Theta(y) \left[ \ln\left( \frac{1+\sqrt{y}}{ 1-\sqrt{y} } \right)-i \pi \right]  
   + 2~\Theta(-y) \arctan\left( \frac{1}{\sqrt{-y}} \right) \right\}.  
\end{align}
The similarity in the $\qsq$ dependence of the $D\Db$, $D\Dstarb$ and $\Dstar\Dstarb$ amplitudes means that the $D\Db$, $D\Dstarb$ and $\Dstar\Dstarb$ two-particle contributions can be effectively approximated with a single term with magnitude $\eta_{\rm 2p}$ and phase $\delta_{\rm 2p}$. This assumption is necessary to ensure fit stability given the data sample size used in this analysis. 

As presented in Ref.~\cite{Cornella:2020aoq}, the two-particle model adopted in this paper allows for the inclusion of a $B^+\to K^+(\tau^+\tau^-\to\mu^+\mu^-)$ nonlocal amplitude whose magnitude is quantified through the lepton-flavour-universality-violating Wilson coefficient $\myC{9\tau}$. However, the approximation of having a single magnitude and phase to describe all three $D\Db$, $D\Dstarb$ and $\Dstar\Dstarb$ amplitudes results in a large systematic uncertainty on a measurement of $\myC{9\tau}$. As such, more data are required to enable the determination of $\myC{9\tau}$ without relying on a simplified treatment of the two-particle charm rescattering amplitudes.

A procedure analogous to that used for the $c\bar{c}$ contributions is adopted for the light-quark nonlocal amplitudes. In this case, an unsubtracted dispersion relation is employed, and the contribution from the light-quark 2p amplitudes is omitted, in line with Ref.~\cite{Cornella:2020aoq}. The resulting light-quark nonlocal contribution is given by
\begin{align}
    Y^{\rm 1p}_{\mathrm{light}}(\qsq) = \sum_{j} \eta_je^{i\delta_j} A^{\rm res}_j (\qsq), 
    \label{eq:light_hadron_amp}
\end{align}
where the sum runs over the \rhossz, \omegaset and \phioztz  states. In this case,  $A_{j}^{\rm res}(\qsq)$ is given by a relativistic Breit--Wigner function shown in Eq.~\ref{eq:relbw}
, while  the \phioztz amplitude,  $A_{j}^{\rm res}(\qsq)$ takes the form defined in  Ref.~\cite{BaBar:2013jqz} that accounts for the open-kaon threshold. The \qsq-dependent width of the  \phioztz amplitude is given by
\begin{align}
\label{eq:phi-updated}
    \Gamma_{\phioztz} (\qsq) = &\left(\frac{p(m_\Kp,\qsq)}{p(m_\Kp,m^2_{\phioztz})}\right)^3 \frac{m_{\phioztz}}{\sqrt{\qsq}}\Gamma_{\decay{\phioztz\,}{\Kp\Km}} \nonumber \\ &+ 
    \left(\frac{p(m_\Kz,\qsq)}{p(m_\Kz,m^2_{\phioztz})}\right)^3 \frac{m_{\phioztz}}{\sqrt{\qsq}}\Gamma_{\decay{\phioztz\,}{\Kz\Kzb}} \nonumber \\ &+ \left (\Gamma_{\phioztz} - \Gamma_{\decay{\phioztz\,}{\Kp\Km}} - \Gamma_{\decay{\phioztz\,}{\Kz\Kzb}} \right),
\end{align}
where $p(m_l,x)$ is given in Eq.~\ref{eq:momentum_eval_atX}, $\Gamma_{\phioztz}$ represents the full decay width of the \phioztz state and $\Gamma_{\decay{\phioztz}{\Kp\Km}}$, $\Gamma_{\decay{\phioztz}{\Kz\Kzb}}$ correspond to the given partial widths. The mass and width of the \phioztz meson are taken from Ref.~\cite{Kozyrev:2017agm}.

The branching fraction of the nonlocal states, $\BF(\decay{\Bp}{\Kp X_j}) \times \BF(\decay{X_j}{\mu^+\mu^-})$, is given by,
\begin{align}
    \frac{\tau_{B}}{\hbar} \frac{G_F^2\alpha^2|\Vtb\Vtss|^2}{128\pi^5} \left| \eta_j\right|^2 \int\limits_{4 m^2_\mu}^{(m_B - m_K)^{2}}|\mathbf{k}|^3 \left[ \beta_+ - \frac{1}{3}\beta^3_+ \right]  \left| f_+(q^2) \right|^2 
     \left| S_j(\qsq)A_j^{\rm res} (\qsq) \right|^2 \deriv q^2,
    \label{eq:reso_bran_frac}
\end{align}
where $\tau_B$ is the $\Bp$-meson lifetime, and $\hbar$ is the reduced Planck's constant.
The factor $S_j(\qsq) = \frac{q^2-q_0^2}{m_j^2-q_0^2}$ for the charmonium resonances and $S_j(\qsq) = 1$ for the light resonances. 
The branching fractions of intermediate 2p states can be calculated by making the substitution $A^{\rm res}_j \rightarrow A^{\rm 2p}_j$ and using $S_j(\qsq) = 1$. 

The parameters of interest in this analysis are the Wilson coefficients $\myC{V}$ and $\myC{A}$, and the magnitudes and phases (\magni{i}, \phase{i}) of the nonlocal amplitudes. The effect of \CP violation has been found to be small across the $m_{\mu\mu}$ range~\cite{LHCb-PAPER-2014-032}, including the region around the \rhossz and \omegaset resonances, whose contributions carry an SM weak phase~\cite{Alok:2011gv}. As a consequence, this analysis is currently not sensitive to \CP-violating effects at low dimuon masses. Therefore, \magni{i} and \phase{i} are taken to be the same for \Bp and \Bm decays throughout.

The magnitudes and phases of all the charmonium resonances are measured from data except for the \jpsi state. 
The \jpsi resonance is used as the normalisation channel and therefore its magnitude is fixed while the phase is measured relative to the local amplitude. 
In Ref. \cite{Cornella:2020aoq} the spectral densities of the 2p states are written using two-body K\"{a}ll\'{e}n functions at their respective thresholds. 
Based on this, the approximate solution to the 2p amplitudes is used in this analysis. 

Two separate computations of the local $B\to K$ form factors are employed in this analysis. The baseline implementation uses the parametrisation and form-factor results of the High Precision Lattice QCD (HPQCD) collaboration~\cite{Parrott:2022rgu}, with those of the Fermilab and MIMD Lattice Computation (FNAL/MILC) collaboration~\cite{Bailey:2015dka} used as a cross-check. In both cases, all form-factor parameters are allowed to vary in the fit to the data, subject to constraints from their respective determinations in Refs.~\cite{Parrott:2022rgu,Bailey:2015dka}.

\section{Detector and simulations}
\label{sec:det_sims}
The \lhcb detector~\cite{LHCb-DP-2008-001,LHCb-DP-2014-002} is a single-arm forward
spectrometer covering the \mbox{pseudorapidity} range $2<\eta <5$,
designed for the study of particles containing \bquark or \cquark
quarks. The detector used to collect the data analysed in this paper includes a high-precision tracking system
consisting of a silicon-strip vertex detector surrounding the $pp$
interaction region~\cite{LHCb-DP-2014-002}, a large-area silicon-strip detector located
upstream of a dipole magnet with a bending power of about
$4{\mathrm{\,T\,m}}$, and three stations of silicon-strip detectors and straw
drift tubes~\cite{LHCb-DP-2014-002} placed downstream of the magnet.
The tracking system provides a measurement of the momentum, \ptot, of charged particles with
a relative uncertainty that varies from 0.5\% at low momentum to 1.0\% at 200\gevc.
The minimum distance of a track to a primary $pp$ collision vertex (PV), the impact parameter (IP), 
is measured with a resolution of $(15+29/\pt)\mum$,
where \pt is the component of the momentum transverse to the beam, in\,\gevc.
Different types of charged hadrons are distinguished using information
from two ring-imaging Cherenkov detectors~\cite{LHCb-DP-2014-002}. 
Photons, electrons and hadrons are identified by a calorimeter system consisting of
scintillating-pad and preshower detectors, an electromagnetic
and a hadronic calorimeter. 
Muons are identified by a
system composed of alternating layers of iron and multiwire
proportional chambers~\cite{LHCb-DP-2012-002}.
The online event selection is performed by a trigger~\cite{LHCb-DP-2014-002}, 
which consists of a hardware stage, based on information from the calorimeter and muon
systems, followed by a software stage, which applies a full event
reconstruction.
Triggered data further undergo a centralised, offline processing step
to deliver physics-analysis-ready data across the entire \lhcb physics programme~\cite{Stripping}.

Simulation is required to model the effects of the detector acceptance, resolution and the
imposed selection requirements.
In the simulation, $pp$ collisions are generated using \pythia~\cite{Sjostrand:2007gs} with a specific \lhcb configuration~\cite{LHCb-PROC-2010-056}.
Decays of unstable particles
are described by \evtgen~\cite{Lange:2001uf}, in which final-state
radiation is generated using \photos~\cite{davidson2015photos}.
The interaction of the generated particles with the detector, and its response,
are implemented using the \geant
toolkit~\cite{Allison:2006ve, *Agostinelli:2002hh} as described in
Ref.~\cite{LHCb-PROC-2011-006}. 
Data driven corrections are applied to simulation to account for the small level of mismodelling of the detector occupancy, \Bp momentum, \Bp vertex quality, and particle identification (PID) performance. 
The momentum of the reconstructed tracks in simulation is also smeared by a small amount to better match the mass resolution of data.

\section{Data Selection}
\label{sec:det_sel}
At first, \BKmumu candidates must pass the hardware stage of the trigger.
In this stage, at least one muon is required to be above a \pt threshold between 1.36\gevc and 1.80\gevc, depending on the run period. 
In the subsequent software stages, at least one of the final state particles is required to have a \pt \greaterthan 1.7\gevc unless it is a muon, in which case, the condition is \pt \greaterthan 1 \gevc. 
The final state particles are also required to have an IP \greaterthan 100 \mum with respect to all PVs to reject particles originating directly from the PV. 
The final stage of the trigger is based on the topology of the signal decay and requires two or more final state particles to form a vertex that is significantly displaced from any PV. 

In the offline stage, a signal candidate is formed by combining a track identified as a kaon with a pair of oppositely charged tracks identified as muons.
The three tracks are required to have a large IP with respect to all PVs and form a good quality vertex. 
For the \Bp candidate, the IP with respect to one of the PVs is required to be small, and the decay vertex is required to be significantly displaced from this PV.
The angle between the reconstructed \Bp momentum and the vector connecting the PV to the reconstructed \Bp decay vertex is required to be small.
The reconstructed \Bp candidate is also required to have mass, $m_{K\mu\mu}$, in the range $4700 (4600) \lessthan m_{K\mu\mu} \lessthan 7000$\mevcc for Run~2 (Run~1). 
To reduce backgrounds arising through the misidentification of final-state particles, PID requirements are placed on variables constructed using information from the RICH, calorimeter, and muon subdetectors for all final-state particles. 

A source of background, commonly referred to as combinatorial, arises from random combinations of kaons and muons from the underlying event, which happen to pass the selections.
To suppress this type of background, a boosted decision tree (BDT) classifier~\cite{BreiFrieStonOlsh84, FREUND1997119} is used. 
Separate BDTs are trained for Run~1 and Run~2 data-taking periods. During the training, both BDTs use simulated \BKmumu decays as the signal proxy, and \BKmumu candidates in the data with $m_{K\mu\mu}$ in the region between 5600 and 5900\mevcc as the background proxy.
To avoid potential biases from resonant dimuon modes in the background proxy, candidates close to the \jpsi and \psitwos resonances are removed. 
A k-folding procedure~\cite{kFold} with k=8 folds is adopted to train the BDT that uses kinematic and geometric variables, including the variables used in the offline selection stage described above. The variables offering the highest discrimination power include the kinematic properties of the \Bp candidate and the difference in the vertex-fit $\chi^2$ of the PV when reconstructed with and without the final state particles. The selection on the BDT criterion is chosen to ensure a signal efficiency of 90\% for both Run~1 and Run~2 data-taking periods, with a background misidentification rate of about 6\% (10\%) for Run~1 (Run~2). 

Background candidates from \B meson decays with a similar topology to \BKmumu are also present in the data.
These originate from pions misidentified as muons from $\decay{\Bp}{\pip (\decay{\Dzb}{\Kp\pim}})$ decays or a kaon-muon swap from $\decay{\Bp}{\Kp(\decay{\jpsi}{\mumu})}$ decays. 
To reduce these to a negligible level, tighter PID criteria in the specific resonant mass windows are employed. The PID requirements on the muons also reduce background contributions from $B^+\to K^+\pi^+\pi^-$ decays to a negligible level. 

\section{Experimental description of the decay rate}
\label{sec:decay_rate_in_data}
To accurately describe the data, the theoretical 
\BKmumu decay rate is extended to account for detector response and background contributions.
This section describes the efficiency model for the event reconstruction and selection, the detector resolution of the dimuon invariant mass, and the modelling of background contributions.

\subsection{Efficiency model}

\label{sec:det_eff}
The reconstruction and selection requirements placed on the signal candidates result in an efficiency variation across the decay phase space completely described through two variables: $\qsq\equiv m_{\mu\mu}^2$ and $\cos\theta_\ell$, where $\theta_\ell$ is the angle between the direction of the \mun(\mup) and the direction of the \Bp(\Bm) meson in the rest frame of the dimuon system. The most significant efficiency-sculpting effects come from the geometrical acceptance of the detector, the IP requirements on the kaon and muon tracks,  and the \pt requirement of the trigger.  The efficiency function, $\mathcal{E}(q^2,\cos\theta_\ell)$, is determined using simulated \BKmumu decays passing the entire selection chain and corrected for differences between data and simulation that are associated with the modelling of the trigger efficiency, PID performance, and \Bp meson kinematic distributions, using control samples in the data. Each of these simulated candidates is assigned a weight according to the phase-space density in \qsq and \ctl.

\begin{figure}
    \centering
    \includegraphics[width=0.7\linewidth]{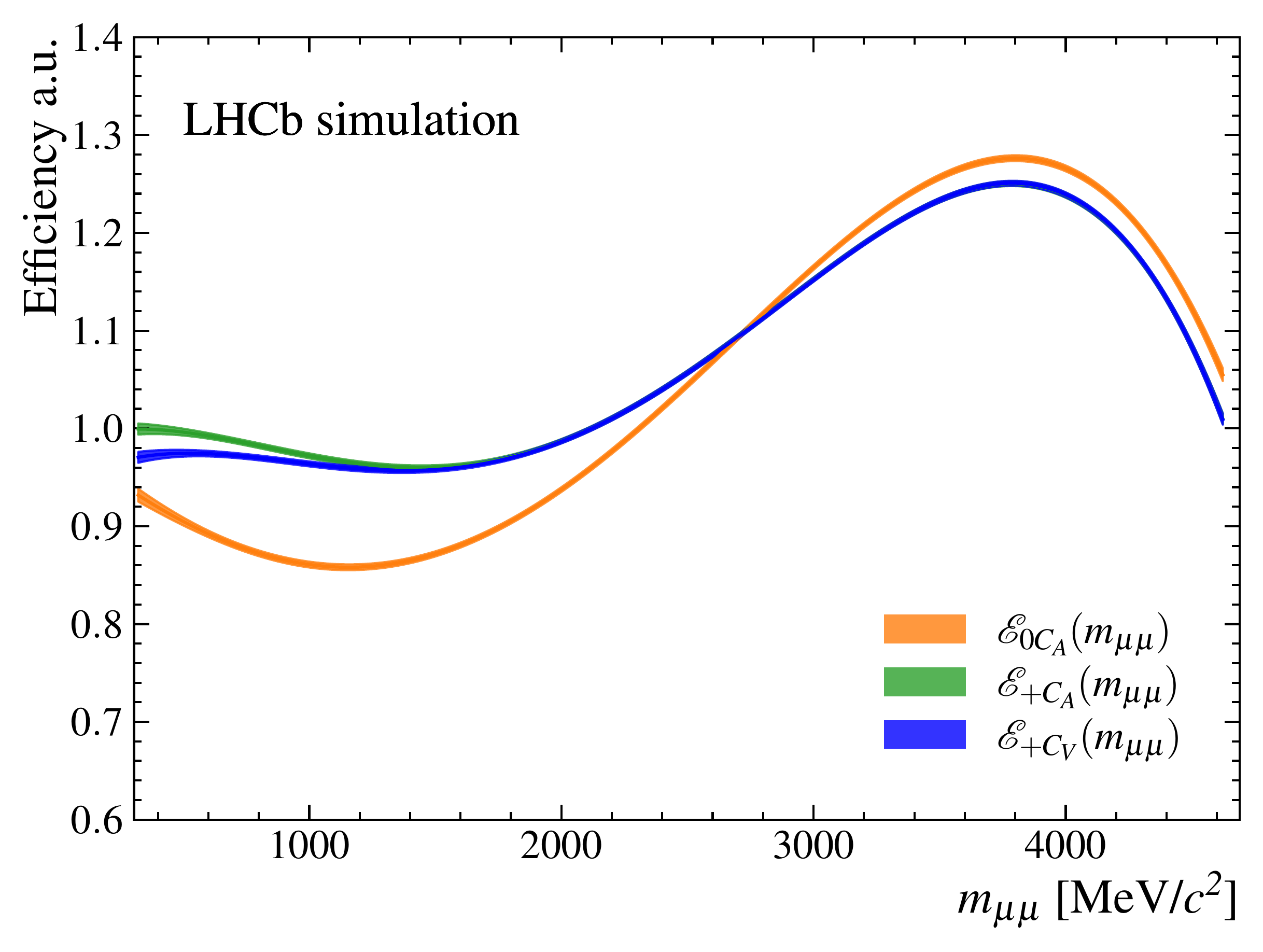}
    \caption{The efficiency functions $\mathcal{E}_{+\myC{A}}$, $\mathcal{E}_{0\myC{A}}$, and $\mathcal{E}_{+\myC{V}}$ given in Eqs.~\ref{eq:eff_fpc10},~\ref{eq:eff_fzc10}, and~\ref{eq:eff_fpc9} respectively. The bands represent the statistical uncertainty due to the size of the simulation sample.}
    \label{fig:efficiency_curve}
\end{figure}

This analysis is performed by integrating over $\cos\theta_\ell$ such that the fully selected \mbox{\BKmumu} candidates are described by the efficiency-corrected decay rate given by
\begin{equation}
    \left. \frac{\deriv\Gamma_\mu}{\deriv q^2}\right |_{\rm eff}\equiv\int\frac{\deriv ^{2}\Gamma_\mu}{\deriv q^2 \deriv \cos\theta_\ell}\mathcal{E}(q^2,\cos\theta_\ell)\deriv \cos\theta_\ell.
    \label{eq:double_diff_bran_frac_int}
\end{equation} 
with 
\begin{align}
    \frac{\deriv^{2}\Gamma_\mu}{\deriv q^2 \deriv \cos\theta_\ell} \propto |\mathbf{k}| \beta_+ & \left \{|\mathbf{k}|^{2}\beta_+ ^2 (1-\cos^2\theta_\ell) |\myC{A} |^2 f_{+}^{2}(q^2) \right.  \nonumber \\ 
    & + \frac{m_{\mu}^2(M_B^2 - M_K^2)^2}{q^{2}M_{B}^{2}} |\myC{A} |^2 f_{0}^{2}(q^2) \nonumber \\ 
    & \left. + |\mathbf{k}|^{2}(1-\beta_+ ^2 \cos^2\theta_\ell) \left | \myCeff{V} + 2 (\myCeff{7}+\myC{7}') \frac{m_b + m_s}{M_B + M_K} \frac{f_T(q^2)}{f_+(q^2)} \right |^2 f_{+}^{2}(q^2) \right \},
    \label{eq:double_diff_bran_frac}
\end{align}
where $\myCeff{V}\equiv \myCeff{9}+\myC{9}'$ and, following Ref.~\cite{Cornella:2020aoq}, we assume no scalar, pseudoscalar, or tensor Wilson coefficients that give rise to an angular term proportional to $\cos\theta_\ell$~\cite{Bobeth:2007dw}. Therefore, from Eqs.~\ref{eq:double_diff_bran_frac_int} and~\ref{eq:double_diff_bran_frac}, the integral of the efficiency-corrected decay rate over $\cos\theta_\ell$ is broken up into three terms with different $\cos\theta_\ell$, Wilson coefficient and form-factor dependence. These terms are given by 
 \begin{alignat}{2}    
     &\mathcal{E}_{+\myC{A}}(\qsq) = \frac{\int^{+1}_{-1} \mathcal{E}(\qsq, \cos\theta_\ell) (1- \cos^2\theta_\ell) \deriv \cos\theta_\ell}{\int^{+1}_{-1} (1- \cos^2\theta_\ell) \deriv \cos\theta_\ell} \label{eq:eff_fpc10},\\ \nonumber
     \\ 
     &\mathcal{E}_{0\myC{A}}(\qsq) = \frac{\int^{+1}_{-1} \mathcal{E}(\qsq, \cos\theta_\ell) \deriv \cos\theta_\ell}{\int^{+1}_{-1}  \deriv \cos\theta_\ell} \label{eq:eff_fzc10},\\ \nonumber
     \\ 
     &\mathcal{E}_{+\myC{V}}(\qsq) =  \frac{\int^{+1}_{-1} \mathcal{E}(\qsq, \cos\theta_\ell) (1-\beta_+ ^2 \cos^2\theta_\ell) \deriv \cos\theta_\ell}{\int^{+1}_{-1} (1-\beta_+ ^2 \cos^2\theta_\ell) \deriv \cos\theta_\ell} \label{eq:eff_fpc9}.
 \end{alignat}
The denominator of these efficiency functions is defined such that the efficiency-corrected differential decay rate, $\left. \frac{\deriv^{2}\Gamma_\mu}{\deriv q^2}\right |_{\rm eff}$, is obtained by multiplying each of the three terms on the right-hand side of Eq.~\ref{eq:diff_bran_frac_theory_ch} by the corresponding efficiency function given in Eqs.~\ref{eq:eff_fpc10},~\ref{eq:eff_fzc10},~\ref{eq:eff_fpc9}.
The efficiencies are determined separately for Run~1 and Run~2 data-taking periods and averaged according to the ratio of yields of fully selected \decay{\Bp}{\Kp(\decay{\jpsi}{\mup\mun})} in each period. The  efficiency distributions defined in  Eqs.~\ref{eq:eff_fpc10},~\ref{eq:eff_fzc10},~\ref{eq:eff_fpc9} are shown in Fig.~\ref{fig:efficiency_curve}. 
Each of these three efficiency distributions is parametrised using a fourth-order polynomial \mbox{in $m_{\mu\mu}$}. 

\subsection{Resolution model}
\label{sec:resolution}
To improve the dimuon-mass resolution, a kinematic fit is performed for each selected \Bp candidate constraining the reconstructed $m_{K\mu\mu}$ to the known \Bp mass~\cite{PDG2024} and requiring the \Bp candidate to originate from one of the PVs in the event. The kinematically constrained dimuon mass is denoted as $m_{\mu\mu}^{\rm rec}$.
After this constraint, the detector resolution varies between 2 and 7\mevcc, depending on dimuon mass, and needs to be taken into account when determining parameters of interest. To account for this, the \BKmumu phase space is split into three $m_{\mu\mu}^{\rm rec}$ regions each containing a dominant resonance as shown in Table~\ref{tab:conv_regions}.  
A sum of a Gaussian function with power-law tails on either side of the peak~\cite{Skwarnicki:1986xj} and a second Gaussian function with the same peak position is used as the resolution model. 
This model is validated on simulation for \jpsi and \psitwos mesons but all the resolution parameters are extracted from data, except for region-1. In this region migration effects from narrow resonances, such as the $\phioztz$, are far smaller compared to those in regions-2 and -3, and the resolution parameters are therefore obtained from simulation.

\begin{table}
    \centering
    \caption[The regions defined for convolution]{The three regions chosen for the convolution and the dominant resonance within the region.}
    \label{tab:conv_regions}
    \begin{tabular}{ccc}
        \toprule
        & $m_{\mu\mu}^{\rm rec}$ range $[\mevcc]$ & Dominant resonance \\
        \midrule
        Region-1 & \phantom{0}300 to 1800 &  \phioztz \\
        Region-2 & 1800 to 3400 & \jpsi \\
        Region-3 & 3400 to 4700 &  \psitwos \\
        \bottomrule
    \end{tabular}
\end{table}

The efficiency-corrected decay rate is convolved with the resolution models to give the experimental description of the signal decay rate as a function of the reconstructed dimuon mass, 
\begin{equation}
    P^{\rm sig}(m^{\rm rec}_{\mu\mu}) =\mathcal{R}(m^{\rm rec}_{\mu\mu}, m_{\mu\mu}) \otimes \left[2m_{\mu\mu} \left. \frac{\deriv\Gamma_\mu}{\deriv q^2}\right |_{\rm eff}  \right].
    \label{eq:conv_dimuon_mass}
\end{equation}
Here $\mathcal{R}(m^{\rm rec}_{\mu\mu}, m_{\mu\mu})$ is one of the three resolution models depending on the region $m^{\rm rec}_{\mu\mu}$.
The factor  $2m_{\mu\mu}$ is the conversion factor from $\deriv q^2$ to $\deriv m_{\mu\mu}$ and $\left.\frac{\deriv\Gamma_\mu}{\deriv q^2}\right |_{\rm eff}$ is the efficiency corrected differential decay rate from Eq. \ref{eq:double_diff_bran_frac_int}.

\subsection{Modelling the background contributions}
\label{sec:bkg}
The residual backgrounds surviving the selection criteria are the combinatorial background and the backgrounds involving pions misidentified as kaons from \mbox{\decay{\Bp}{\pip\mup\mun}} decays.
The yields of these backgrounds are estimated from unbinned maximum-likelihood fits to the reconstructed $m_{K\mu\mu}$ mass. 
The model of the $m_{K\mu\mu}$ distribution depends on the dimuon mass distribution. Therefore, the signal and background yields are extracted in the three regions of dimuon mass shown in Table~\ref{tab:conv_regions}.

\begin{figure}
    \centering
    \includegraphics[width=0.49\textwidth]{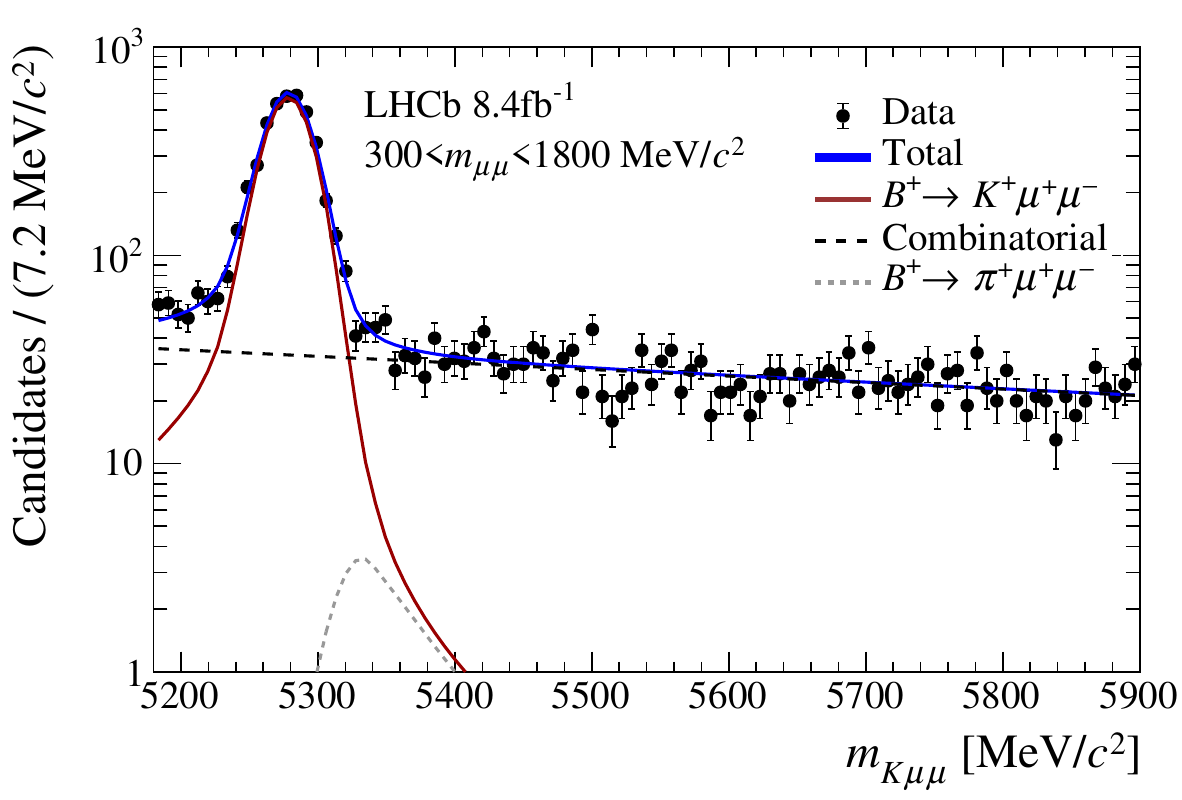}
    \includegraphics[width=0.49\textwidth]{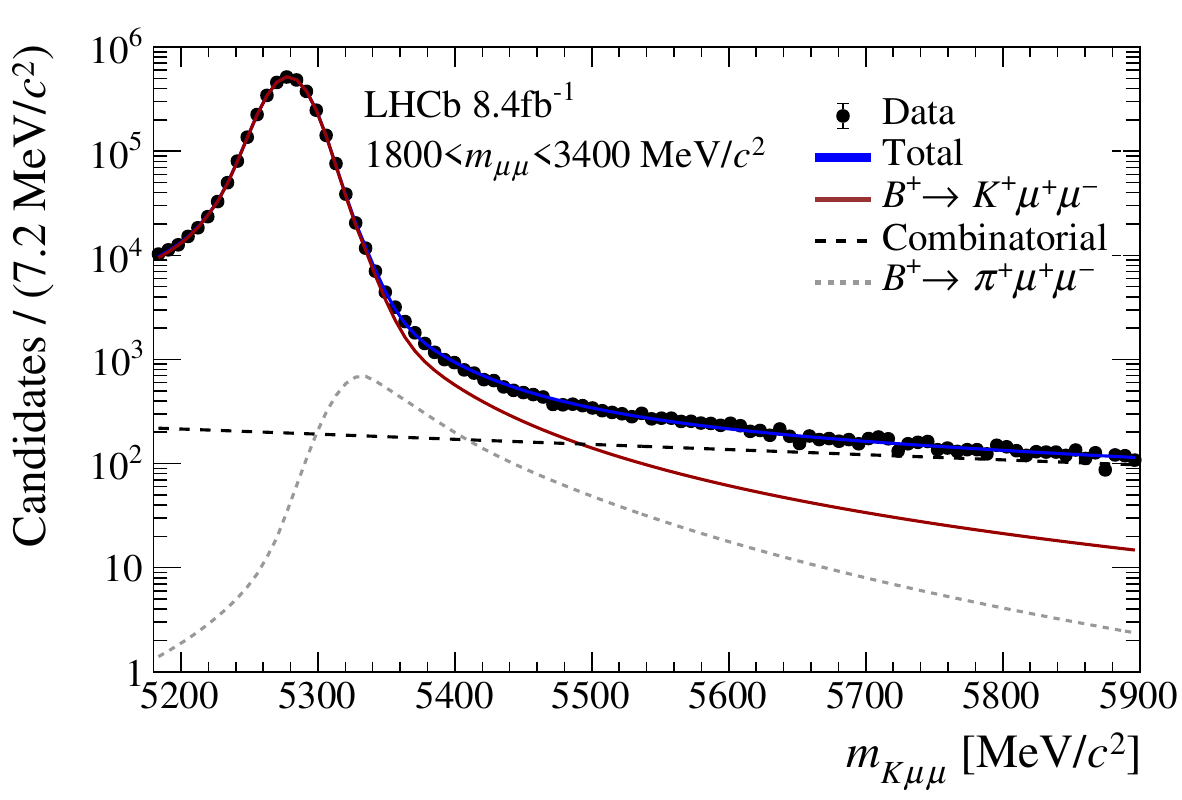}\\
    \includegraphics[width=0.49\textwidth]{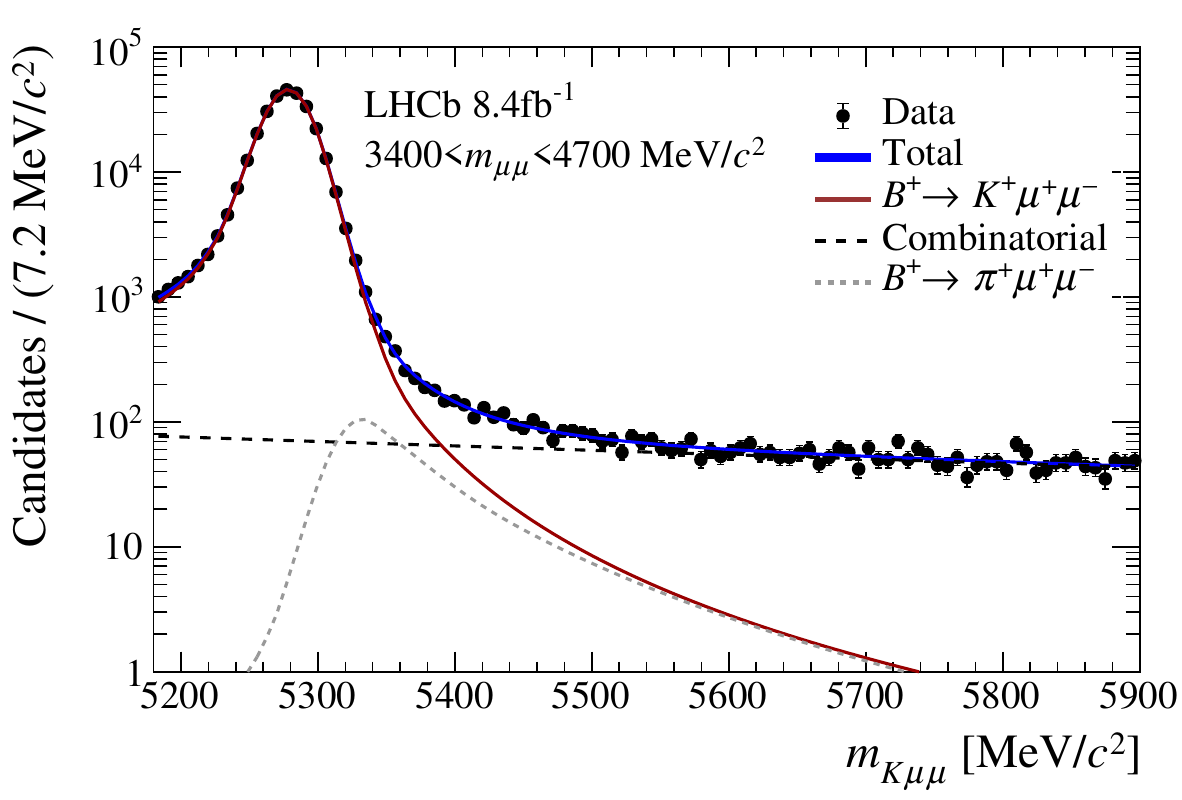}
    \caption{The fit to the reconstructed $m_{K\mu\mu}$ mass distribution in (top left)  region-1, (top right)  region-2, and (bottom) region-3 corresponding to the $m_{\mu\mu}$ ranges defined in Table~\ref{tab:conv_regions}.
    \label{fig:Bmass_peak_fit_results}}
\end{figure}

The reconstructed $m_{K\mu\mu}$ distribution for the three dimuon mass regions are shown in Fig.~\ref{fig:Bmass_peak_fit_results}.
An exponential function is used to model the combinatorial background and a Gaussian distribution with power law tails~\cite{Skwarnicki:1986xj} is used to model the misidentified pion background, whose shape parameters are extracted from simulated \decay{\Bp}{\pip\mup\mun} candidates. The yield of \decay{\Bp}{\pip\mup\mun} decays is constrained in the fit using the central value and uncertainty of the measured \decay{\Bp}{\pip\mup\mun} branching fraction presented in Ref.~\cite{LHCB-PAPER-2015-035}.
The signal shape in $m_{K\mu\mu}$ is modelled using a Gaussian function with power-law tails added and two other Gaussian functions all with shared peak positions. 
The peak position and width parameters are determined from the fit to the data while the tail parameters are fixed to values determined from simulation.

The fit to the $m_{\mu\mu}^{\rm rec}$ distribution is performed for candidates with $m_{K\mu\mu}$ within a signal window of $\pm40\mevcc$ of the known $B^+$ mass. The fits to the $m_{K\mu\mu}$ distribution in the data are used to determine the signal and background fractions within this window. Kernel density estimators (KDEs)~\cite{Cranmer:2000du} are used to model the $m_{\mu\mu}$ shapes of misidentified $\decay{\Bp}{\pip(\decay{\jpsi}{\mup\mun})}$ and $\decay{\Bp}{\pip(\decay{\psitwos}{\mup\mun})}$ decays in the simulation. The contribution of misidentified pions from the decay of $\decay{\Bp}{\pip\mup\mun}$ within the $m_{K\mu\mu}$ signal window is negligible and is therefore ignored.

Combinatorial \BKmumu candidates comprise three components: a combinatorial $K^+$ with a genuine \jpsi candidate (combinatorial \jpsi); a combinatorial $\Kp$ with a genuine \psitwos candidate (combinatorial \psitwos); and a three-particle fully-combinatorial $\Kp\mu^+\mu^-$ component (fully combinatorial). 
The shape of each combinatorial background contribution in $m_{\mu\mu}^{\rm rec}$ is obtained from the $m_{K\mu\mu}$ upper-mass sideband defined as \mbox{$5440 (5460) \lessthan m_{K\mu\mu} \lessthan 5840 (5860)\mevcc$} for the Run~2 (Run~1) data-taking periods. These regions are dominated by combinatorial background candidates with small contributions from $\decay{\Bp}{\Kp\mup\mun}$ and $\decay{\Bp}{\pip\mup\mun}$ decays. The kinematic constraint introduces a dependence between the $m_{K\mu\mu}$ distribution and the kinematically constrained $m_{\mu\mu}^{\rm rec}$ distribution. Therefore, in order to model this dependence, the upper-mass sideband is split into five 80\mevcc-wide $m_{K\mu\mu}$ regions with the kinematic constraint applied to the centre of each region, and the dimuon mass is recomputed using this kinematic constraint. 
The kinematically constrained $m_{\mu\mu}^{\rm rec}$ data in each of these five regions is fitted using unbinned maximum-likelihood estimators. An ARGUS function~\cite{ARGUS:1994rms} is used to model the continuum combinatorial background, the sum of two Gaussian functions for the \jpsi combinatorial background, and a single  Gaussian function for \psitwos combinatorial background. 
The combinatorial background parameters describing the $m_{\mu\mu}^{\rm rec}$ shape include the ARGUS shape parameter, the masses and widths of the Gaussians describing the \jpsi and \psitwos contributions, and the fraction between the two Gaussians used to model the \jpsi combinatorial component.
The $m_{\mu\mu}^{\rm rec}$ distribution of \decay{\Bp}{\Kp\mup\mun} and \decay{\Bp}{\pip\mup\mun} decays misreconstructed in the upper-mass sideband are modelled using KDEs determined from simulation in each of the \mKmumu upper-mass sideband regions. The yields of these misreconstructed components are constrained to their expected values from the fit to the \mKmumu distribution in the data. 
These parameters and the fraction of combinatorial \jpsi and \psitwos over the total combinatorial background are extrapolated linearly to the known \Bp mass giving the shape of the combinatorial background. 
The shapes of the background from the two run periods are added together weighted by the background yields in the signal region.
The total model is given by the sum of the signal and background components, where the signal fraction is obtained from the fits to the \mKmumu distribution in each of the three $m_{\mu\mu}^{\rm rec}$ regions.

\section{Systematic uncertainties}
\label{sec:syst}

The lattice QCD precision in the local $B\to K$ form factors is an important source of systematic uncertainty. These form-factor uncertainties are propagated into the likelihood through an external multivariate Gaussian constraint on the form-factor coefficients, provided by the different lattice QCD calculations~\cite{Parrott:2022rgu,Bailey:2015dka}. Similarly, the uncertainties on the masses and widths of the broad resonances above the open-charm threshold are propagated into the likelihood through Gaussian constraints provided in Ref.~\cite{BES:2007zwq}.

Another important source of systematic uncertainty originates from the precision of the measured branching fraction of \decay{\Bp}{\Kp\jpsi} decays as taken from Ref.~\cite{Jung:2015yma}. The size of this contribution is evaluated by repeating the fit with the branching fraction of the \decay{\Bp}{\Kp\jpsi(\to\mu^+\mu^-)} decay varied according to its uncertainty. The RMS of the resulting distribution of each parameter of interest, along with the linear correlations of these variations between the parameters, comprise the covariance matrix of the systematic uncertainty.

The effect of using three distinct resolution functions in the fit instead of a single resolution function with a continuous dependence on the dimuon mass is estimated using pseudoexperiments. Each pseudoexperiment is generated using a model based on the best fit point in the data, without any convolution applied, and then smeared using two different sets of resolution models: one that accounts for the continuous dependence of the resolution on dimuon mass, and another following the baseline procedure of three separate resolution functions in each of the three dimuon mass regions. Both sets of pseudodata are then fitted with the baseline fit model. The systematic covariance matrix of the parameters of interest is evaluated by taking the RMS width of the distribution resulting from the differences between the two fit results for each parameter of interest, along with the linear correlations in between the parameter variations.

The statistical precision of the background model is propagated to the parameters of interest by bootstrapping the upper-mass sideband data, varying the signal fractions according to their covariances in each dimuon mass region, and repeating the background fit procedure outlined in Sec.~\ref{sec:bkg}. The RMS and bias of the fitted parameter distribution are combined in quadrature, and the linear correlations among parameter variations are used to construct the covariance matrix of this systematic uncertainty.

The uncertainty on the subtraction constant $Y^0_\ccbar=0.230\pm0.065$~\cite{Bordone:2024hui} is propagated directly to \myC{9} and is entirely uncorrelated with other parameters of interest. Other systematic uncertainties that were studied and found to be negligible include the statistical precision of the efficiency model, the choice of the efficiency parametrisation, and the impact of the corrections applied to the simulation. 

Table~\ref{tab:all_systematics} summarises the size of the sources of systematic uncertainty described in this section. The systematic uncertainty from the precision of the $B\to K$ form factors determined by the HPQCD collaboration~\cite{Parrott:2022rgu} is indicative and directly encoded in the likelihood surface. The total systematic uncertainty is obtained by adding up the systematic covariance matrices and smearing the one- and two-dimensional likelihood profiles presented in Sec.~\ref{sec:result_and_discussion}.

\begin{table}[!tb]
    \centering
\caption{Systematic uncertainties on the parameters of interest. See text for further information. The systematic uncertainty associated with the precision of the HPQCD $B\to K$ form factors are indicative, as this uncertainty is directly encoded in the likelihood.}
    \label{tab:all_systematics}
    \scalebox{0.9}{
\begin{tabular}{lccccc}
\toprule
Parameter & HPQCD FF & $\mathcal{B}(B^+\to \jpsi K^+)$ &  Resolution & Bkg extrapolation & Subtraction \\
\midrule
    $\myC{V}$      & 0.141 & 0.079 &   0.013 & 0.012 & 0.065 \\
    $\myC{A}$     & 0.084 & 0.014 &  0.034 & 0.031 & -- \\ 
\midrule
    \phase{\rhossz}~~[rad] &  0.074  & 0.002 & 0.016     & 0.005 & -- \\ 
    \magni{\rhossz}      & 0.012 & 0.001 & 0.001     & 0.001 & -- \\  
\midrule
    \phase{\omegaset}~~[rad] & 0.006 & 0.015 &  0.018     & 0.001 & -- \\  
    \magni{\omegaset}   & 0.058 & 0.005 &  0.009     & 0.001 & -- \\ 
\midrule
    \phase{\phioztz}~[rad]  & 0.005 & 0.051 &  0.004     & 0.002 & -- \\      
    \magni{\phioztz}     & 0.082 & 0.004 & 0.010     & 0.001 & -- \\ 
\midrule                                  
    \phase{\jpsi}~~~~~[rad]      &0.018 & 0.001 &  0.002 & 0.006 & -- \\ 
\midrule         
                                         
    \phase{\psitwos}\hspace{0.09in}[rad]  & 0.027 & 0.006 &  0.006 & 0.012 & -- \\  
    \magni{\psitwos}    & 41.42 & 18.69 &  0.020 & 0.087 & -- \\ 
\midrule        
    \phase{\psitss}~[rad]  &  0.030 & 0.002 &  0.003 & 0.011 & -- \\   
    \magni{\psitss}     & 0.099 & 0.035 &  0.006 & 0.040 & -- \\ 
\midrule     
                      
    \phase{\psifzfz}~[rad] & 0.098 & 0.003 &  0.006 & 0.018 & -- \\ 
    \magni{\psifzfz}    & 0.135 & 0.015 &  0.009 & 0.006 & -- \\ 
\midrule       
                   
    \phase{\psifosz}~[rad] &  0.107 & 0.002 &  0.007 & 0.009 & -- \\ 
\magni{\psifosz}  &  0.118 & 0.023 &  0.013 & 0.010 & -- \\ 
\midrule  
                                     
    \phase{\psiffof}~[rad] &  0.085 & 0.002 &  0.016 & 0.010 & -- \\ 
    \magni{\psiffof}   & 0.072 & 0.016 &  0.027 & 0.009 & -- \\ 
\midrule      
    \phase{\rm 2p}~~~~~~[rad] & 0.207 & 0.003 &  0.035 & 0.017 & -- \\     
    \magni{\rm 2p}   & 0.137 & 0.002 &  0.005 & 0.021 & -- \\ 
\bottomrule
\end{tabular}
}
\end{table}

\section{Results and discussion}
\label{sec:result_and_discussion}
An unbinned log-likelihood fit is performed to the dimuon spectrum of \BKmumu candidates using the signal and background models described in Sec.~\ref{sec:decay_rate_in_data}. The parameters of interest of the fit are the Wilson coefficients $\myC{V}$, $\myC{A}$, and the magnitudes and phases of the nonlocal amplitudes. The baseline fit relies on the most recent and precise lattice QCD determinations of the $B^+\to K^+$ local form factors provided by the HPQCD collaboration~\cite{Parrott:2022rgu}, while an alternative fit using local $B\to K$ form factors by the FNAL/MILC collaboration~\cite{Bailey:2015dka} is also explored. 

\begin{figure}[tb]
    \centering
    \includegraphics[width=0.49\linewidth]{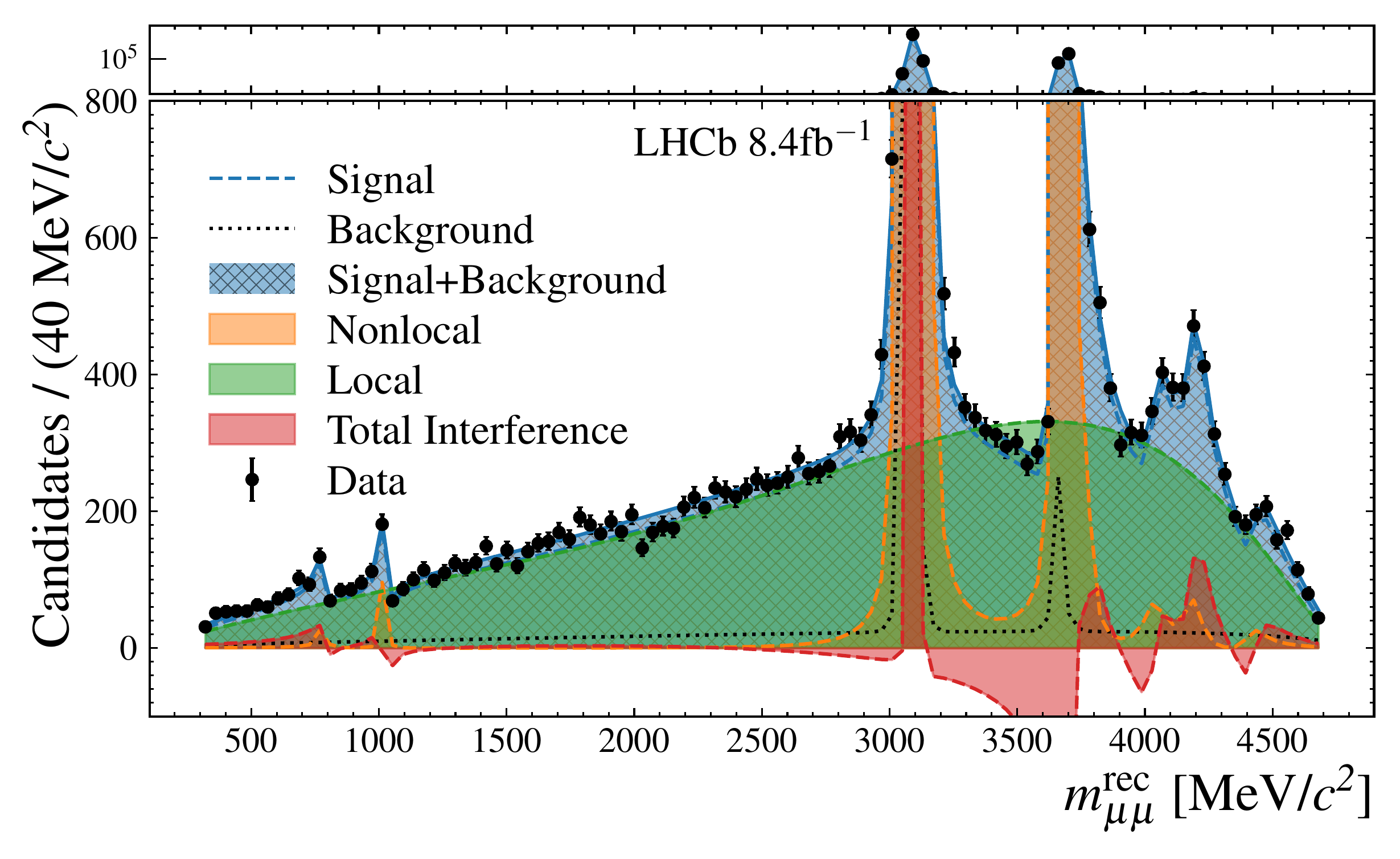}
    \includegraphics[width=0.49\linewidth]{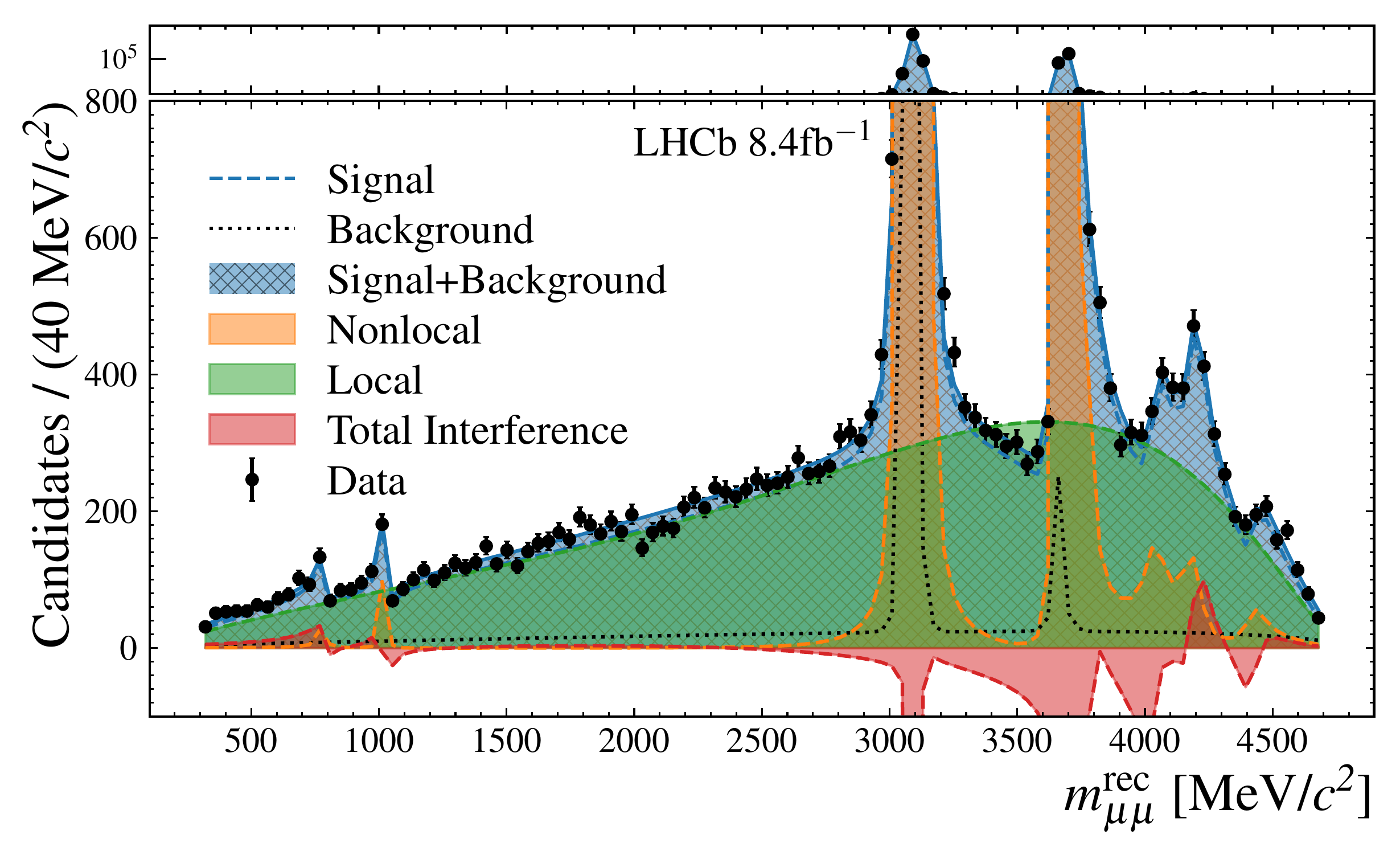}\\
    \includegraphics[width=0.49\linewidth]{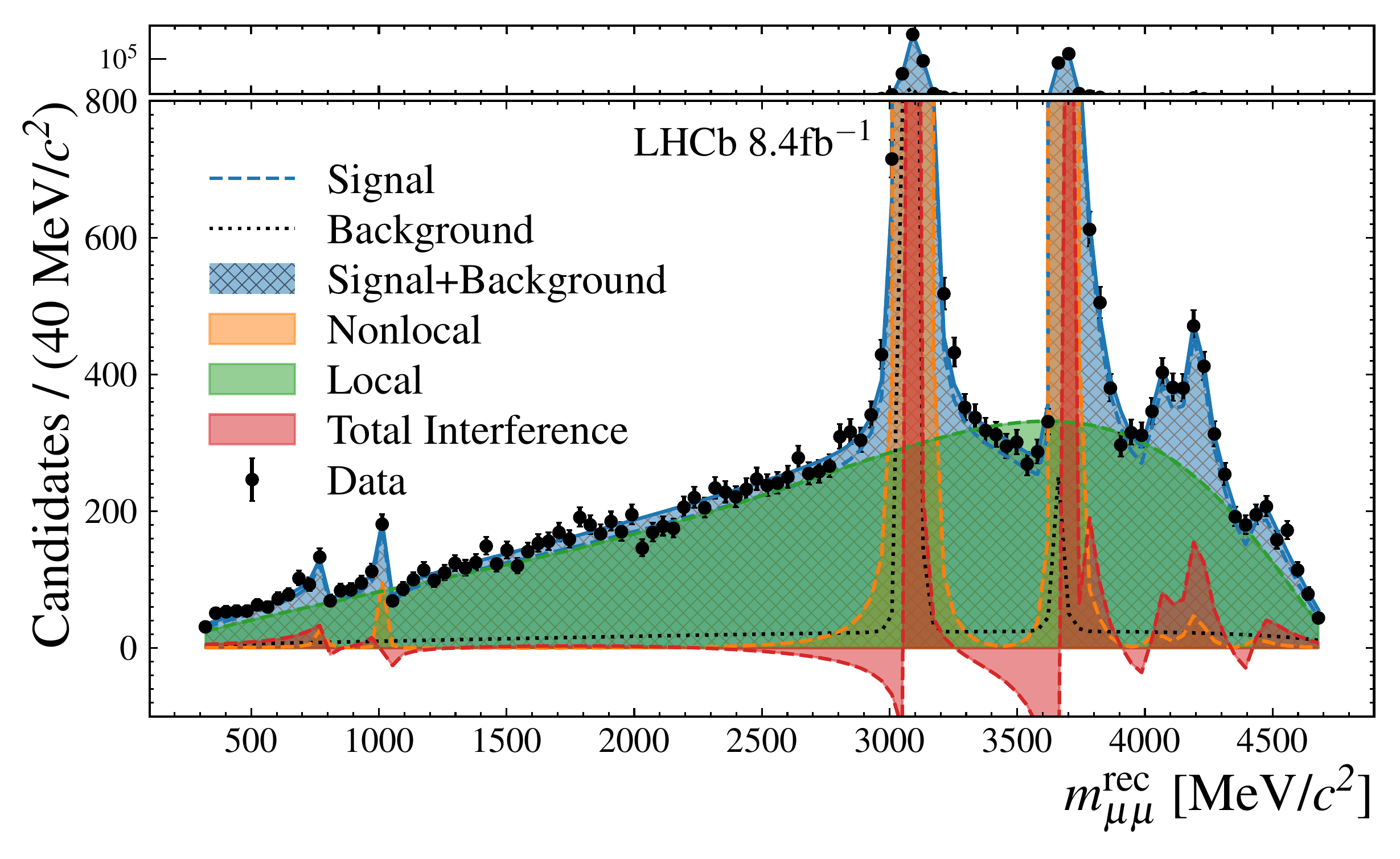}
    \includegraphics[width=0.49\linewidth]{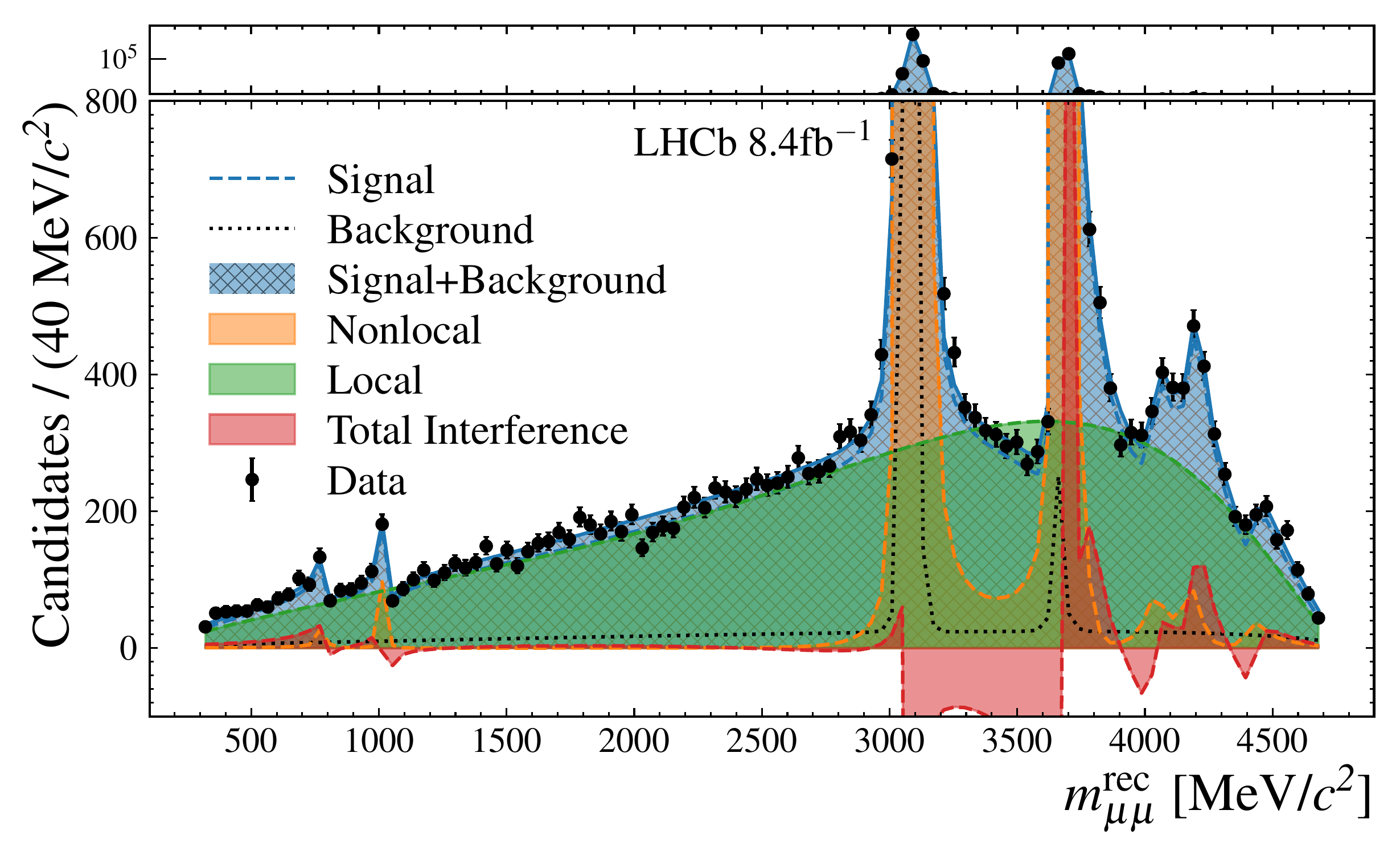}
    \caption{Fit to the $m^{\rm rec}_{\mu\mu}$  distribution using form factors from the HPQCD collaboration~\cite{Parrott:2022rgu} for: (top left) negative~\jpsi phase, positive~\psitwos phase; (top right) positive~\jpsi phase, positive~\psitwos phase; (bottom left) negative~\jpsi phase, negative~\psitwos phase; (bottom right) positive~\jpsi phase, negative~\psitwos phase.}
    \label{fig:HPQCD_dimuon_fit}
\end{figure}

Similarly to what was reported in the previous iteration of this analysis~\cite{LHCb-PAPER-2016-045}, the negative log-likelihood exhibits four near-degenerate minima depending on the signs of the \jpsi and \psitwos resonance phases. The $m_{\mu\mu}^{\rm rec}$ distribution of \BKmumu decays, along with the best-fit model to the data for all four near-degenerate minima, is shown in Fig.~\ref{fig:HPQCD_dimuon_fit}.

The solution with the lowest negative log-likelihood is found for a negative $J/\psi$ phase and a positive $\psi(2S)$ phase relative to the local amplitude, with a small destructive quantum-mechanical interference in the decay rate. 
Table~\ref{tab:4-fold_deg_summary_min2pconfig} presents the changes in the negative log-likelihood and Wilson coefficients $\myC{V}$ and $\myC{A}$ in all four quadrants relative to the quadrant with the lowest negative log-likelihood. The Wilson coefficients remain essentially identical for all four solutions. In contrast, the nonlocal parameters, presented in Table~\ref{tab:non-local-results}, vary between the solutions. Similarly to the measurement of the nonlocal amplitudes in $B^0\to K^{*0}\mu^+\mu^-$ decays presented in Ref.~\cite{LHCb-PAPER-2024-011}, there is no evidence of two-particle nonlocal contributions from 
$B^+\to(\DorDstar\DorDstarb\to\mu^+\mu^-)K^+$ transitions in the spectrum of $B^+\to K^+\mu^+\mu^-$ decays. 

As discussed in Sec.~\ref{sec:diff_rate}, the number of \BKmumu candidates available in the Run~1 and Run~2 datasets of the LHCb experiment does not allow for the determination of all three two-particle amplitudes. Instead, an effective amplitude is used, in which all three $D\Db$, $D\Dstarb$, and $\Dstar\Dstarb$ components share the same magnitude and phase. Although this choice has a negligible impact on the couplings $\myC{V}$ and  $\myC{A}$, it hinders an accurate determination of $\myC{9}^{\tau}$ entering through the two-particle nonlocal amplitude of the process $B^+\to (\tau^{+}\tau^{-}\to\mu^+\mu^-)K^+$.

\begin{table}[!tb]
    \centering
    \caption{Changes to the negative log-likelihood and Wilson coefficients relative to the deepest minimum for all four-fold degenerate solutions in $\delta_\jpsi$ and $\delta_\psitwos$ phases. The four fit results are denoted as  $(-,+)$, $(+,+)$, $(-,-)$, $(+,-)$, where the first (second) sign in each pair represents the sign of $\delta_\jpsi$ ($\delta_\psitwos$) phase. The values of $\myC{V}$ and $\myC{A}$ are calculated relative to the result in the lowest negative log-likelihood phase quadrant $(-,+)$.}\label{tab:4-fold_deg_summary_min2pconfig}
\begin{tabular}{lcccc}
    \toprule
    Parameter & $(-,+)$ & $(+,+)$ & $(-,-)$ & $(+, -)$\\    
    \midrule
    $-2\Delta\log{\mathcal{L}}^{\rm min}$ & 0 & 0.09  & 0.69  & 0.463 \\
    $|\Delta\myC{V}|$                      & 0 & 0.006 & 0.006 & 0.014 \\ 
    $|\Delta\myC{A}|$                      & 0 & 0.001 & 0.011 & 0.012 \\
    \bottomrule
\end{tabular}
\end{table}

\begin{table}[!t]
    \centering
    \caption{Measured values of the nonlocal parameters in the different (\phase{\jpsi}, \phase{\psitwos}) quadrants. The four fit results are denoted as  $(-,+)$, $(+,+)$, $(-,-)$, $(+,-)$ where the first (second) sign in each pair represents the sign of $\delta_\jpsi$ ($\delta_\psitwos$) phase. The first uncertainty is statistical while the second is systematic.}
\label{tab:non-local-results}

\scalebox{0.85}{
\begin{tabular}{lcccc}
    \toprule
    Parameter & $(-, +)$ & $(+, +)$ & $(-,-)$ & $(+, -)$\\    
    \midrule
    \phase{\rhossz}~~[rad]             &  $-0.48\pm0.36\pm0.02$& $-0.47\pm0.36\pm0.02$ &  $-0.49\pm0.36\pm0.02$    &  $-0.48\pm0.36\pm0.02$   \\  
    \magni{\rhossz}                &  $\phantom{+}0.69\pm0.06\pm0.00$ & $\phantom{+}0.69\pm0.07\pm0.00$  & $\phantom{+}0.69\pm0.06\pm0.00$    & $\phantom{+}0.69\pm0.06\pm0.00$    \\  
    \midrule
    
    \phase{\omegaset}~~[rad]        &  $-0.08\pm0.26\pm0.02$ & $-0.07\pm0.26\pm0.02$  & $-0.08\pm0.26\pm0.02$    & $-0.07\pm0.26\pm0.02$    \\   
    \magni{\omegaset}             &  $\phantom{+}4.65\pm0.29\pm0.01$ & $\phantom{+}4.64\pm0.36\pm0.01$  & \phantom{+}$4.65\pm0.31\pm0.01$    &  $\phantom{+}4.65\pm0.32\pm0.01$    \\   
    \midrule
    
    \phase{\phioztz}~[rad]     &  $\phantom{+}0.38\pm0.31\pm0.05$ & $\phantom{+}0.40\pm0.31\pm0.05$  & $\phantom{+}0.38\pm0.30\pm0.05$    &  $\phantom{+}0.40\pm0.30\pm0.05$  \\        
    \magni{\phioztz}                    & $12.77\pm0.87\pm0.01$ & $12.77\pm0.96\pm0.01$  & $12.77\pm0.87\pm0.01$    & $12.77\pm0.91\pm0.01$   \\ 
    \midrule                                          
    \phase{\jpsi}~~~~~[rad]     &  $-1.56\pm0.06\pm0.01$ & $\phantom{+}1.60\pm0.05\pm0.01$  &  $-1.70\pm0.07\pm0.01$    & $\phantom{+}1.45\pm0.04\pm0.01$     \\ \midrule

    \phase{\psitwos}\hspace{0.09in}[rad]     & $\phantom{+}2.21\pm0.16\pm0.01$ & $\phantom{+}1.77\pm0.13\pm0.01$ & $-2.02\pm0.13\Spm0.01$    & $-2.46\pm0.11\pm0.01$  \\    
    \magni{\psitwos} \hspace{0.07in}$(\times 10^3)$                 & $\phantom{+}1.23 \pm 0.03 \pm 0.02$ & $\phantom{+}1.23 \pm 0.03 \pm 0.02$ & $\phantom{+}1.23 \pm 0.03 \pm 0.03$ & $\phantom{+}1.24 \pm 0.04 \pm 0.02$ \\       

    \midrule      
    
    \phase{\psitss}~[rad]     &  $\phantom{+}2.90\pm0.17\pm0.01$ & $\phantom{+}2.50\pm0.35\pm0.01$& $-2.56\pm0.18\pm0.01$    &  $-2.96\pm0.13\pm0.01$ \\        
    \magni{\psitss}                       &  $\phantom{+}2.92\pm0.43\pm0.05$ & $\phantom{+}2.93\Spm0.68\pm0.05$ & $\phantom{+}2.68\pm0.40\pm0.05$    & $\phantom{+}2.72\pm0.31\pm0.05$ \\    
    \midrule      
    
    \phase{\psifzfz}~[rad]     & $-3.11\pm0.19\pm0.02$ & $\phantom{+}2.86\pm0.32\pm0.02$ & $-2.82\pm0.18\pm0.02$ & $-3.14\pm0.17\pm0.02$      \\  
    \magni{\psifzfz}                 & $\phantom{+}1.50\pm0.26\pm0.02$ & $\phantom{+}1.52\pm0.30\pm0.02$  & $\phantom{+}1.45\pm0.25\pm0.02$ & $\phantom{+}1.46\pm0.25\pm0.02$\\ 
    \midrule

    \phase{\psifosz}~[rad]     & $-2.31\pm0.18\pm0.01$ & $-2.60\pm0.32\pm0.01$  & $-2.08\pm0.17\pm0.01$ & $-2.37\pm0.18\Spm0.01$   \\   
    \magni{\psifosz}                 & $\phantom{+}2.04\pm0.26\pm0.03$ & $\phantom{+}2.05\pm0.32\pm0.03$  & $\phantom{+}2.02\pm0.25\pm0.03$ & $\phantom{+}2.03\pm0.24\pm0.03$\\
    \midrule

    \phase{\psiffof}~[rad]      & $\phantom{+}2.56\pm0.32\pm0.02$ & $\phantom{+}2.32\pm0.47\pm0.02$ & $\phantom{+}2.70\pm0.30\pm0.02$ & $\phantom{+}2.45\Spm0.25\pm0.02$  \\  
    $\magni{\psiffof}$               & $\phantom{+}1.62\pm0.32\pm0.03$ & $\phantom{+}1.64\pm0.37\pm0.03$ & $\phantom{+}1.59\pm0.33\pm0.03$ & $\phantom{+}1.61\pm0.34\pm0.03$  \\
    \midrule          
    $\phase{\rm 2p}$~~~~~~[rad]      & $-2.41\pm0.56\pm0.04$ & $-2.63\pm0.79\pm0.04$  & $-2.29\pm0.60\pm0.04$ & $-2.49\pm0.47\pm0.04$   \\        
    $\magni{\rm 2p}$              & $\phantom{+}0.30\pm0.16\pm0.02$ & $\phantom{+}0.28\pm0.20\pm0.02$ & $\phantom{+}0.32\pm0.20\pm0.02$ & $\phantom{+}0.31\pm0.14\pm0.02$   \\
    \bottomrule
\end{tabular}
}
\end{table}

\subsection{Measurement of Wilson coefficients}
\begin{figure}[!h]
    \centering
    \includegraphics[width=0.49\linewidth]{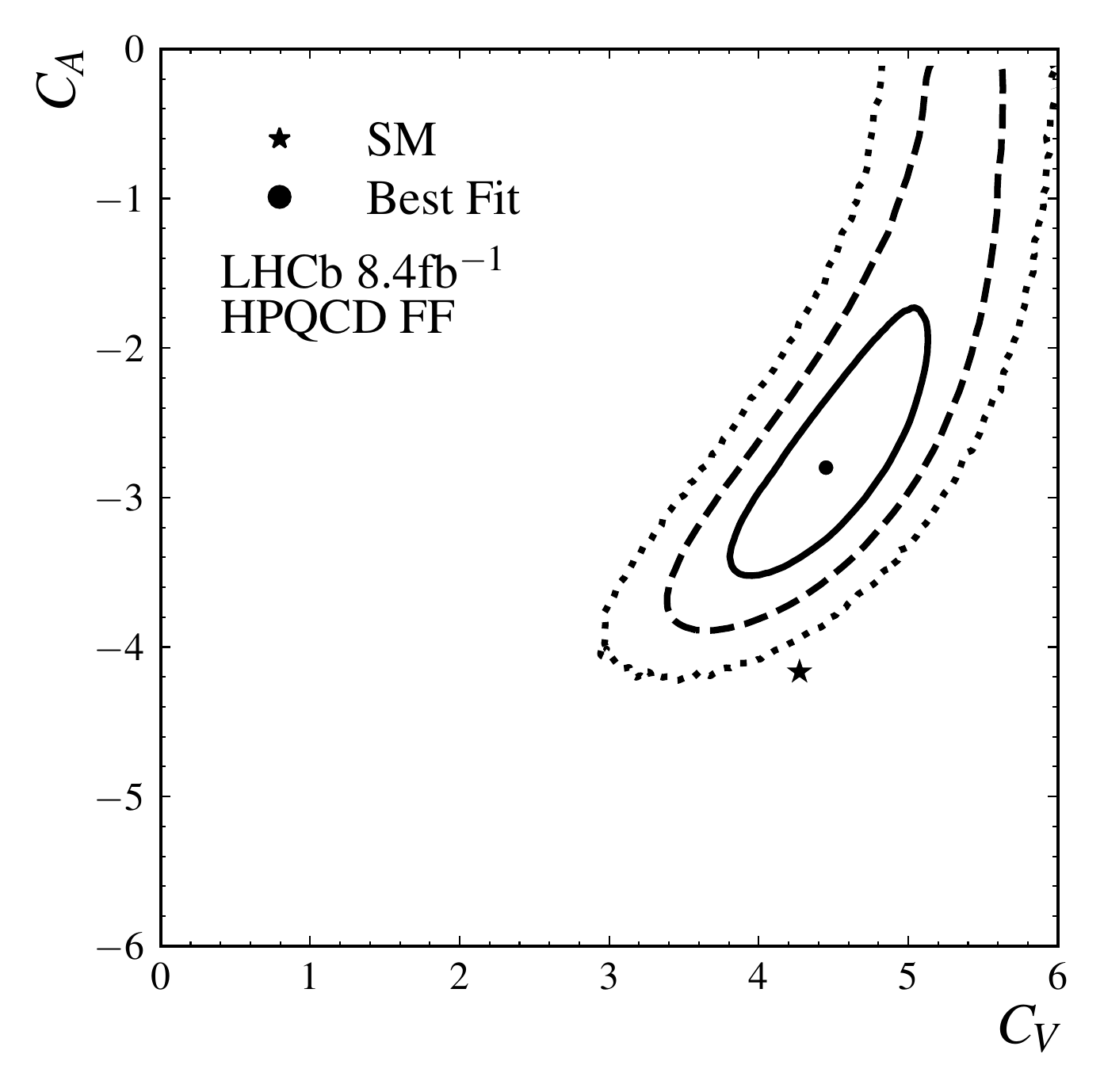}
    \includegraphics[width=0.49\linewidth]{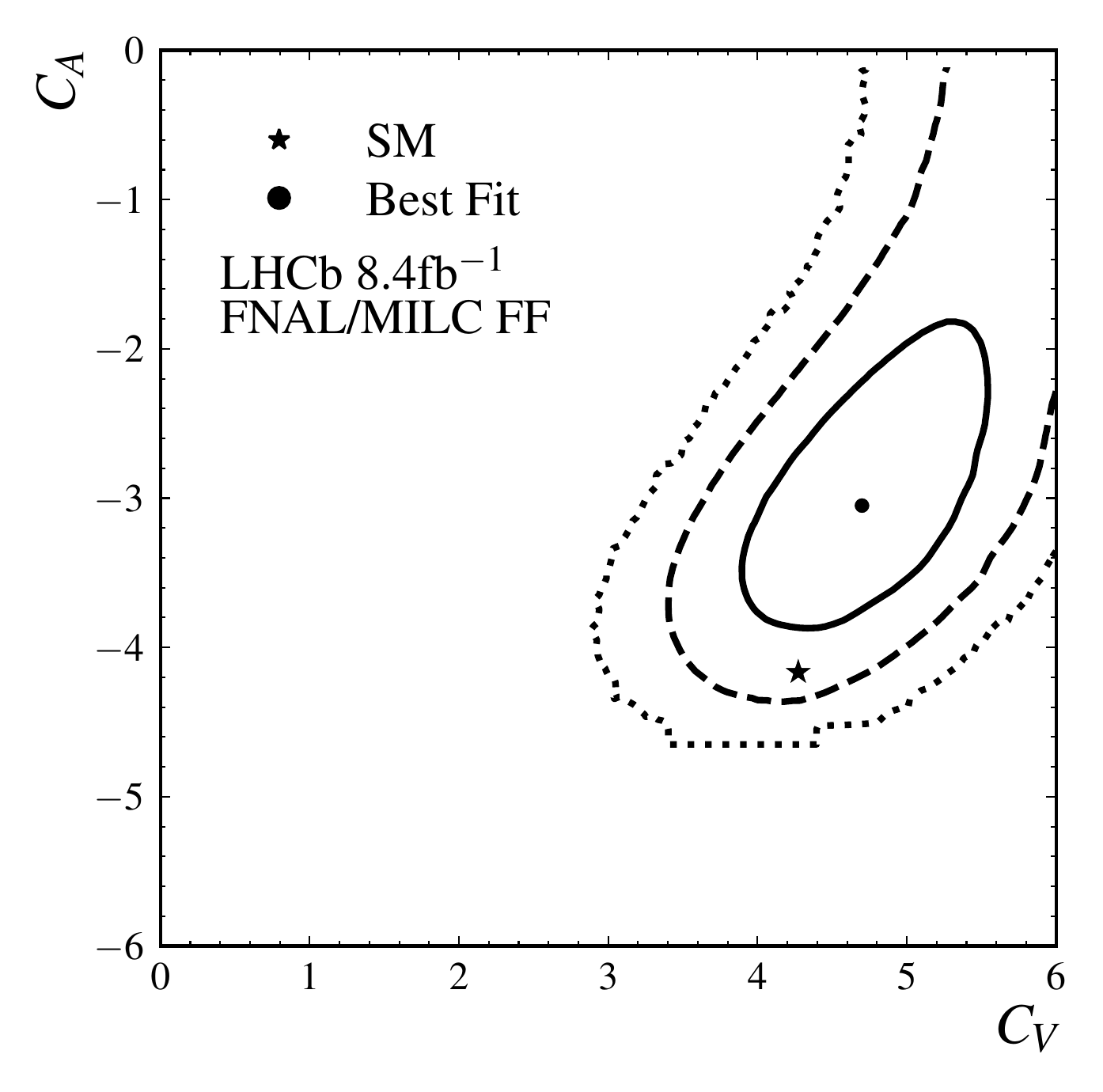}
    \caption{Likelihood profiles of \myC{V} and \myC{A} resulting from a fit to the data using the $B^+\to K^+$ local form-factor determination from (left) the HPQCD collaboration~\cite{Parrott:2022rgu} and (right) the FNAL/MILC collaboration~\cite{Bailey:2015dka}. The different curves correspond to the 68.3\%, 95.4\% and 99.7\% confidence intervals that include systematic uncertainties. The SM point is shown as a star.}
    \label{fig:HPQCD-2D-Wilson}
\end{figure}
The two-dimensional likelihood surface for \myC{V} and \myC{A}, including all systematic uncertainties, is shown in Fig.~\ref{fig:HPQCD-2D-Wilson}. The branching fraction of the short-distance component provides a strong constraint on the combination $\myC{V}^2+\myC{A}^2$, as shown in Eq.~\ref{eq:diff_bran_frac_theory_ch}, resulting in the annular structure of the likelihood profile in Fig.~\ref{fig:HPQCD-2D-Wilson}. The fit distinguishes between $\myC{V}$ and $\myC{A}$ through the interference of $\myC{V}$ with the nonlocal amplitudes. For comparison, the corresponding determination obtained using the FNAL/MILC $B \to K$ local form factors from Ref.~\cite{Bailey:2015dka} is also displayed. When employing the HPQCD local form factors for $B\to K$ transitions, the extracted Wilson coefficients exhibit a tension with the Standard Model at the level of \smhpqcdsig. The results obtained using the FNAL/MILC form factors are fully consistent with those derived from the HPQCD calculation, although as expected, the larger uncertainties associated with the FNAL/MILC form factors reduce the significance of the deviation from the Standard Model prediction to \smfnalsig.

\begin{figure}[!h]
    \centering
    \includegraphics[width=0.49\linewidth]{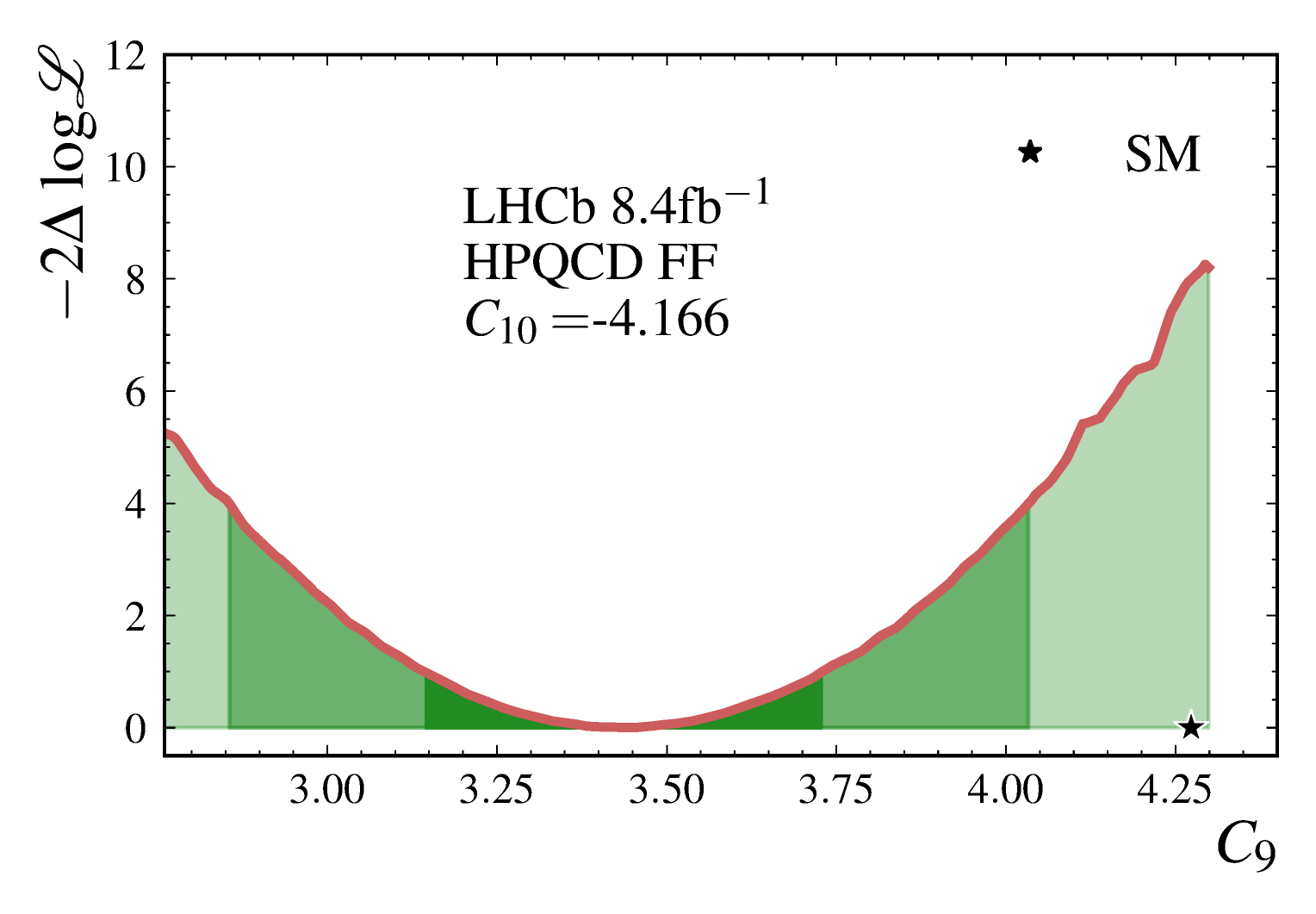}
    \includegraphics[width=0.49\linewidth]{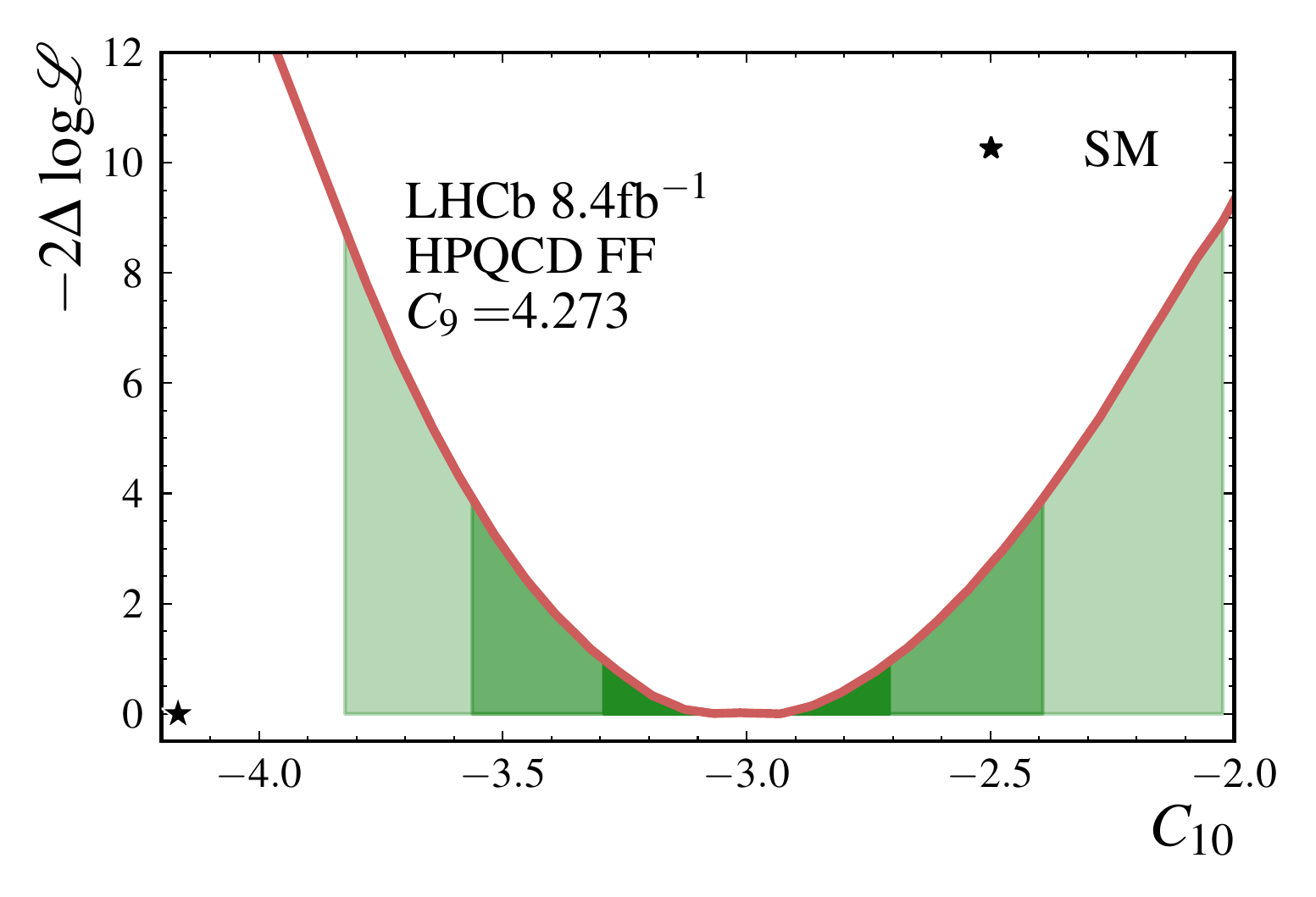}
    \caption{Likelihood profile of (left) $\myC{9}$ while fixing $\myC{10}$ to the SM value, and (right) $\myC{10}$ while fixing $\myC{9}$ to the SM value. In both cases, $\myC{9}'$ and $\myC{10}'$ are set to their SM value of zero, and the HPQCD local form-factor computations are used. The 68.3\%, 95.4\% and 99.7\% confidence intervals are indicated as green bands and include systematic uncertainties. The SM point (star) is also shown.}
    \label{fig:HPQCD-FNAL-1D-Wilson-SM}
\end{figure}

For both HPQCD and FNAL/MILC form factors, the fit to the data prefers values for \myC{V,A} larger than $\myC{V,A}^{\rm SM}$. Assuming SM values for $\myCp{9}=\myCp{10}=0$ and $\myC{10}=-4.166$, the data prefer $\myC{9}<\myC{9}^{\rm SM}$, as shown by the likelihood profiles of Fig.~\ref{fig:HPQCD-FNAL-1D-Wilson-SM}. The one-dimensional profile of $\myC{9}$ obtained using HPQCD local form factors yields a value consistent with global analyses of \bsmumu transitions that indicate potential new physics contributions to \myC{9} (\eg Ref.~\cite{Gubernari:2022hxn}). The observed shift, however, is smaller than that reported in Ref.~\cite{Bordone:2024hui}, based solely on \BKmumu differential branching fraction measurements from Run~1 of the LHCb experiment and Run~2 of the CMS experiment.

\subsection{The differential branching fraction}
\begin{figure}
    \centering
    \includegraphics[width=0.9\linewidth]{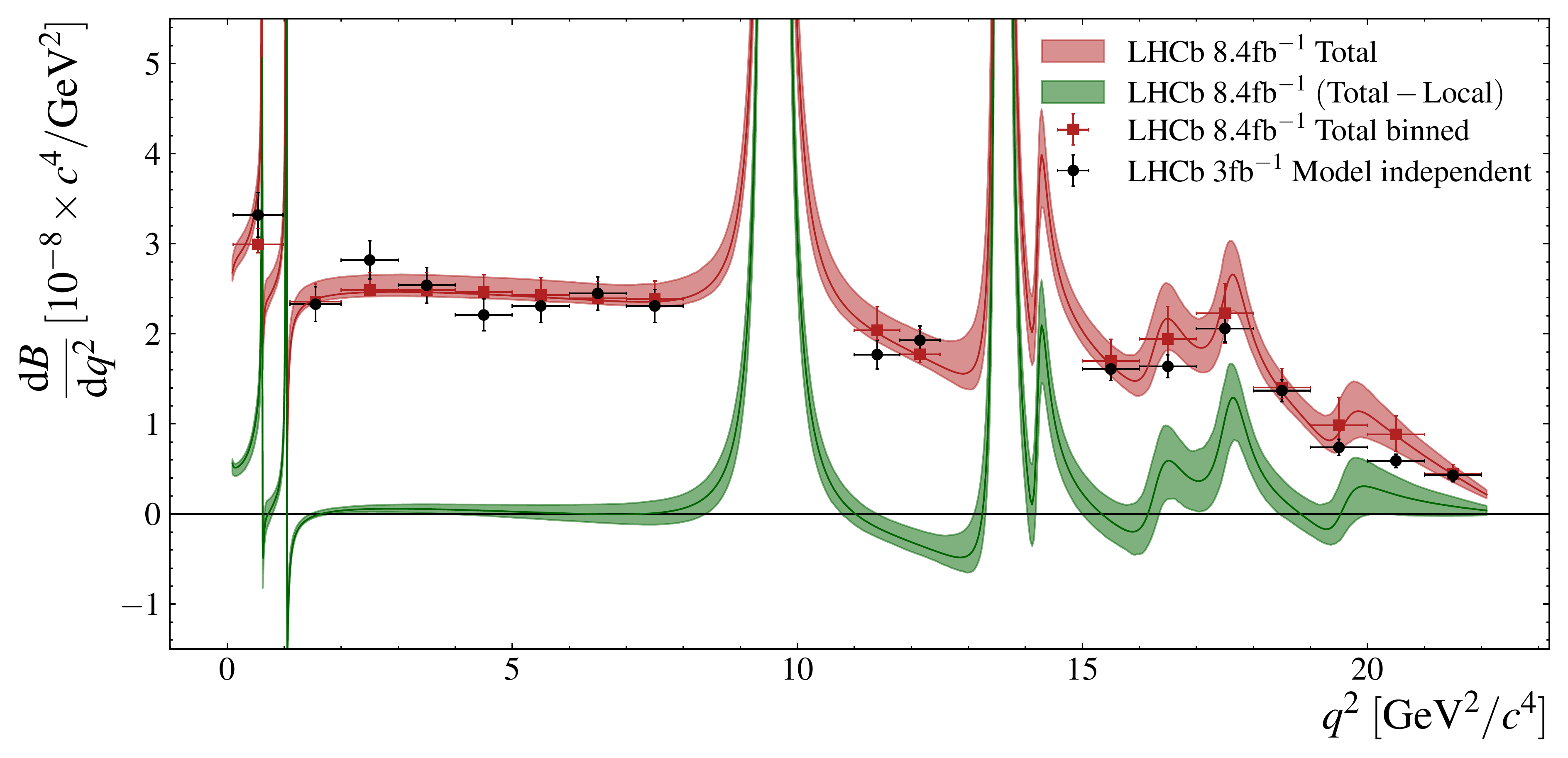}
    \caption{Measured differential branching fraction of the \BKmumu decay shown using the model described in this paper in an unbinned manner (maroon solid curve) and evaluated in \qsq bins (maroon squares with error bars). The contribution of the nonlocal amplitude (green), obtained by subtracting the decay rate associated with the local contribution from the total decay rate, is also displayed. The shaded bands represent the 68.3\% confidence intervals derived from variations of the fit parameters according to the full covariance matrix, incorporating both statistical and systematic uncertainties. For comparison, the model-independent \qsq-binned measurement reported by the LHCb collaboration in Ref.~\cite{LHCb-PAPER-2014-006} (black circles) is also shown.} 
    \label{fig:run1_diff_bran_frac_comp}
\end{figure}

The differential branching fraction is constructed out of the local and nonlocal amplitudes and compared to the direct measurement from the LHCb collaboration using 3\invfb of Run~1 data\cite{LHCb-PAPER-2014-006} as shown in Fig.~\ref{fig:run1_diff_bran_frac_comp}.
A good agreement is seen between the two determinations. Figure~\ref{fig:run1_diff_bran_frac_comp} also shows the effect of the nonlocal amplitude and its interference with the local part in the \BKmumu decay rate. In the region
$1.5 \lessthan\ \qsq\lessthan 8.0 \gevgevcccc$ the nonlocal effects are minor, but become important outside this region.

\subsection{Introducing a $\boldsymbol{\qsq}$-dependent term to Wilson coefficients  $\boldsymbol{\myC{V}}$ and $\boldsymbol{\myC{A}}$}
To check for any residual \qsq-dependent nonlocal contribution to \myC{V} and \myC{A}, alternative fits with \qsq-dependent Wilson coefficients are performed. 
For these fits the Wilson coefficients are reparametrised with a linear dependence on \qsq as
\begin{align}
    \myC{V}^{\qsq} = \myC{V} + a (\qsq - 11.09\gevgevcccc), \quad \myC{A}^{\qsq} = \myC{A} + b (\qsq - 11.09\gevgevcccc),
\end{align}
with $a$ and $b$ being the newly introduced fit parameters.
The point 11.09\gevgevcccc is chosen as it is the midpoint of the considered \qsq phase space. This fit is performed in two variations, where either $a$ or $b$ is treated as a fit parameter. In addition, a step-like dependence on \myC{V} is also investigated and given by the ansatz,
\begin{align}
    \myC{V}^{\rm step} = \myC{V} + \Delta \left( \frac{e^x}{1+e^x} - \frac{1}{2}\right)\mathrm{ with }\quad x=\frac{\mmumu-m_{t}}{100\mevcc}
\end{align}
where $\Delta$ is the newly introduced fit parameter and $m_t=3096.65$\mevcc (\jpsi peak), or $m_t=3871.70$\mevc ($D\Dstarb$ threshold).
The results of the fits introducing a linear \qsq dependence to \myC{V} or \myC{A} are \mbox{$a=0.078\pm0.052\,(\!\gevgevcccc)^{-1}$} and \mbox{$b=-0.090\pm0.045\,(\!\gevgevcccc)^{-1}$}. The introduction of a step-like behaviour in \myC{V} with a threshold at 3096.65\mevcc results in \mbox{$\Delta_{\jpsi}=0.41\pm0.24$}, and with a threshold at 3871.70\mevcc in \mbox{$\Delta_{D\Dstarb}=-1.37\pm0.53$}.

Given the precision of the $\qsq$-dependent parameters, there is a preference for a residual $\qsq$ dependence in \myC{A} at the level of $2\,\sigma$. The value and significance of this parameter are similar to those obtained by the corresponding amplitude analysis in $B^0\to K^{*0}\mu^+\mu^-$ decays presented in Ref.~\cite{LHCb-PAPER-2024-011}.  Alternatively, there is an indication of a threshold effect in \myC{V} above the open-charm threshold at a significance of $2.6\,\sigma$. The systematic uncertainties described in Sec.~\ref{sec:syst} have a negligible effect on these significances.

\section{Conclusions}
\label{sec:conclusion}
Using data collected with the LHCb detector corresponding to an integrated luminosity of 8.4\invfb, an analysis of the local and nonlocal contributions to the 
\BKmumu decay rate has been performed. The nonlocal amplitudes are modelled with a dispersion-relation approach that incorporates both one-particle and two-particle contributions. This represents a significant improvement over the treatment of nonlocal effects in previous LHCb analyses of these decays.

Using the latest lattice QCD computations of the $B\to K$ local form factors from the HPQCD collaboration, the values of the couplings $\myC{V}$ and $\myC{A}$ are found to be in tension with the SM prediction at the \smhpqcdsig level. Under the assumption of no new physics contributions to $\myC{9}'$ and $\myC{10}'$ and $\myC{10}$, the measured value of \myC{9} is compatible with global analyses of \bsmumu transitions~\cite{Gubernari:2022hxn}, although the observed shift is smaller than that reported in Ref.~\cite{Bordone:2024hui}. As reported in Ref.~\cite{LHCb-PAPER-2016-045}, the nonlocal parameters exhibit a four-fold ambiguity that depends on the relative phases of the \jpsi and \psitwos amplitudes. Consequently, all four solutions for the nonlocal parameters are reported.

A fit to \BKmumu candidates has also been performed using previous lattice QCD computations of the $B\to K$ local form factors from the FNAL/MILC collaboration. In this case, the tension with the SM prediction of the couplings $\myC{V}$ and $\myC{A}$ is reduced to \smfnalsig owing to the larger uncertainties of the local form-factor computations. 

When using HPQCD form factors, there is a mild preference for either a residual \qsq dependence in $\myC{A}$, or a threshold effect above the open-charm threshold in \myC{V}. These features, although not statistically significant, may indicate missing amplitude components in the physics model or limitations in its ability to describe the data. The reconstructed  dimuon mass distribution used to perform this analysis, along with the background, efficiency and resolution models, are provided in digital form in the \textsc{Zenodo} page~\cite{lhcb_collaboration_2026_19497184} to facilitate further studies. Additional data from Run~3 of the LHC will be essential to further improve the overall accuracy and precision of the measurement.

%% file: acknowledgements.tex
\section*{Acknowledgements}
%
%
\noindent We express our gratitude to our colleagues in the CERN
accelerator departments for the excellent performance of the LHC. We
thank the technical and administrative staff at the LHCb
institutes.
We acknowledge support from CERN and from the national agencies:
ARC (Australia);
CAPES, CNPq, FAPERJ and FINEP (Brazil); 
MOST and NSFC (China); 
CNRS/IN2P3 and CEA (France);  
BMFTR, DFG and MPG (Germany);
INFN (Italy); 
NWO (Netherlands); 
MNiSW and NCN (Poland); 
MCID/IFA (Romania); 
MICIU and AEI (Spain);
SNSF and SER (Switzerland); 
NASU (Ukraine); 
STFC (United Kingdom); 
DOE NP and NSF (USA).
We acknowledge the computing resources that are provided by ARDC (Australia), 
CBPF (Brazil),
CERN, 
IHEP and LZU (China),
IN2P3 (France), 
KIT and DESY (Germany), 
INFN (Italy), 
SURF (Netherlands),
Polish WLCG (Poland),
IFIN-HH (Romania), 
PIC (Spain), CSCS (Switzerland), 
GridPP (United Kingdom),
and NSF (USA).  
We are indebted to the communities behind the multiple open-source
software packages on which we depend.
Individual groups or members have received support from
Key Research Program of Frontier Sciences of CAS, CAS PIFI, CAS CCEPP (China); 
Minciencias (Colombia);
EPLANET, Marie Sk\l{}odowska-Curie Actions, ERC and NextGenerationEU (European Union);
A*MIDEX, ANR, IPhU and Labex P2IO, and R\'{e}gion Auvergne-Rh\^{o}ne-Alpes (France);
Alexander-von-Humboldt Foundation (Germany);
ICSC (Italy); 
Severo Ochoa and Mar\'ia de Maeztu Units of Excellence, GVA, XuntaGal, GENCAT, InTalent-Inditex and Prog.~Atracci\'on Talento CM (Spain);
the Leverhulme Trust, the Royal Society and UKRI (United Kingdom).

%% file: Authorship_LHCb-PAPER-2025-055.tex
\centerline
{\large\bf LHCb collaboration}
\begin
{flushleft}
\small
R.~Aaij$^{38}$\lhcborcid{0000-0003-0533-1952},
A.S.W.~Abdelmotteleb$^{58}$\lhcborcid{0000-0001-7905-0542},
C.~Abellan~Beteta$^{52}$\lhcborcid{0009-0009-0869-6798},
F.~Abudin\'en$^{60}$\lhcborcid{0000-0002-6737-3528},
T.~Ackernley$^{62}$\lhcborcid{0000-0002-5951-3498},
A. A. ~Adefisoye$^{70}$\lhcborcid{0000-0003-2448-1550},
B.~Adeva$^{48}$\lhcborcid{0000-0001-9756-3712},
M.~Adinolfi$^{56}$\lhcborcid{0000-0002-1326-1264},
P.~Adlarson$^{86}$\lhcborcid{0000-0001-6280-3851},
C.~Agapopoulou$^{14}$\lhcborcid{0000-0002-2368-0147}
% ...

{\footnotesize \it

% ...
$^{14}$Universit{\'e} Paris-Saclay, CNRS/IN2P3, IJCLab, Orsay, France\\
% ...
$^{38}$Nikhef National Institute for Subatomic Physics, Amsterdam, Netherlands\\
% ...
$^{48}$Instituto Galego de F{\'\i}sica de Altas Enerx{\'\i}as (IGFAE), Universidade de Santiago de Compostela, Santiago de Compostela, Spain\\
% ...
$^{52}$Physik-Institut, Universit{\"a}t Z{\"u}rich, Z{\"u}rich, Switzerland\\
% ...
$^{56}$H.H. Wills Physics Laboratory, University of Bristol, Bristol, United Kingdom\\
% ...
$^{58}$Department of Physics, University of Warwick, Coventry, United Kingdom\\
% ...
$^{60}$School of Physics and Astronomy, University of Edinburgh, Edinburgh, United Kingdom\\
$^{61}$School of Physics and Astronomy, University of Glasgow, Glasgow, United Kingdom\\
$^{62}$Oliver Lodge Laboratory, University of Liverpool, Liverpool, United Kingdom\\
% ...
$^{70}$Syracuse University, Syracuse, NY, United States\\
% ...
$^{86}$Department of Physics and Astronomy, Uppsala University, Uppsala, Sweden, associated to $^{61}$\\
% ...
}
\end{flushleft}